%% file: article.tex
\documentclass[useAMS,usenatbib,usegraphicx]{mn2e}

\usepackage{amsmath,amssymb,amsbsy,amsfonts}
\usepackage[british]{babel}
\usepackage{mdwmath}
\usepackage{url}
\usepackage{relsize}
\usepackage{graphicx}
\usepackage{mathrsfs}
\usepackage{units}
\usepackage{hyperref}

\newcommand{\ud}{\mathrm{d}}
\newcommand{\sn}{\ensuremath{\mathrm{sn}}}
\newcommand{\cn}{\ensuremath{\mathrm{cn}}}
\newcommand{\dn}{\ensuremath{\mathrm{dn}}}
\newcommand{\cd}{\ensuremath{\mathrm{cd}}}
\newcommand{\cnsn}{\ensuremath{\mathrm{cs}}}

\newcommand{\sech}{\ensuremath{\mathrm{sech}}}
\newcommand{\s}[1]{{\text{\tiny $#1$ }}\hspace{-3pt}}

\renewcommand{\theequation}{\thesection.\arabic{equation}}


\title[Toy model for relativistic accretion]{An analytic toy model for
relativistic accretion in Kerr spacetime}
\author[E.~Tejeda, P.~A.~Taylor \& J.~C.~Miller]
       {Emilio Tejeda$^1$\footnotemark[1], 
	Paul A. Taylor$^{2}$\footnotemark[1] 
	and John C. Miller$^{3,1}$\thanks{E-mail: tejeda@sissa.it (ET);  
	ptaylor@astro.ox.ac.uk (PT); jcm@astro.ox.ac.uk (JCM).} \\
        $^1$ SISSA \& INFN, Via Bonomea 265, 34136, Trieste, Italy \\
        $^2$ Department of Radiology, UMDNJ-New Jersey Medical 
        School, ADMC 5, Suite 575, 30 Bergen St., 
        Newark, NJ 07103, USA\\
        $^3$ Department of Physics (Astrophysics), 
        University of Oxford, Keble Road, Oxford OX1 3RH, UK }

\pagerange{\pageref{firstpage}--\pageref{lastpage}}

\begin{document}

\maketitle

\label{firstpage}

\begin{abstract}
We present a relativistic model for the stationary axisymmetric accretion flow
of a rotating cloud of non-interacting particles falling onto a Kerr black hole.
Based on a ballistic approximation, streamlines are described analytically in
terms of timelike geodesics, while a simple numerical scheme is introduced for
calculating the density field. A novel approach is presented for describing all
of the possible types of orbit by means of a single analytic expression. This
model is a useful tool for highlighting purely relativistic signatures in the
accretion flow dynamics coming from a strong gravitational field with
frame-dragging. In particular, we explore the coupling due to this between the
spin of the black hole and the angular momentum of the infalling matter.
Moreover, we demonstrate how this analytic solution may be used for benchmarking
general relativistic numerical hydrodynamics codes by comparing it against
results of smoothed particle hydrodynamics simulations for a collapsar-like
setup. These simulations are performed first for a ballistic flow (with zero
pressure) and then for a hydrodynamical one where we measure the effects of
pressure gradients on the infall, thus exploring the extent of applicability of
the ballistic approximation. 
\end{abstract}


\section{Introduction}
\label{introduction}

Matter falling down the potential well of a gravitating object is the
fundamental mechanism behind some of the most powerful astrophysical phenomena
in the universe \citep[see e.g.][]{frank02}. The more compact the central
object, the deeper into the well the matter can reach, and so the greater the
quantity of potential energy available for extraction. However, for black holes
(BHs), the most compact objects in the universe, purely radial infall is
inefficient at converting kinetic energy into radiation since there is no resisting
surface at which decelerate the infalling gas \citep{shapiro74}. Traditionally,
rotation of the accreting matter has been invoked (and also observed) as the
means for providing, at least temporarily, centrifugal support to give time for
different dissipative processes to take place and release part of the binding
energy in the form of radiation. Gas rotation can lead to the formation of a
disc-like structure \citep{pb68} and, indeed, accretion discs around BHs are the
most commonly studied engines for explaining astrophysical phenomena such as
active galactic nuclei \citep{genzel10}, X-ray binaries \citep{king95} and
gamma-ray bursts (GRBs) \citep{piran04}, and they may be produced as a possible
outcome of tidal disruptions \citep{rosswog09} and binary coalescence
\citep{lee10}. Comprehensive analyses of these systems require full-scale,
numerical magnetohydrodynamic simulations in a curved spacetime, together with
an accurate microphysical description of the dissipative processes, such as
cooling and shock formation. Nevertheless, typically two ingredients play the
leading roles in determining the overall accretion efficiency: the gravitational
field generated by the central BH and rotation of the fluid. An examination of
the combination of these last two features is the focus of the present study.

Regarding the gravitational field, it has also been recognised that, in any
realistic scenario, the central BH will posses some amount of intrinsic angular
momentum, either because it was born with it or as a result of accretion of
matter with large angular momentum \citep{bardeen70,blanford87}. Therefore, for
applications in which it is safe to neglect the surrounding mass-energy
contribution to the overall spacetime curvature, the exterior metric around a
physical BH will be well approximated by the Kerr solution. Any substantial
value for the angular momentum of the BH will significantly affect the innermost
region of an accretion disc around the BH, exactly where one expects to find the
highest densities, temperatures and luminosities. For instance, the inner radius
of a Keplerian-like accretion disc around a maximally rotating BH is around six
times closer to the central accretor than it would be for a non-rotating BH
\citep[see e.g.][]{novikov73}, while the binding energy for the innermost stable
circular orbit increases from $\sim 5.7\%$ of the rest-mass energy for a
non-rotating BH up to $\sim 42\%$ for a maximally rotating one. This increase
in both the surface area of the emitting disc and the binding energy released
can substantially boost the overall efficiency of the system. Moreover, the BH
angular momentum may play a relevant role in launching and accelerating a jet
via e.g. the Blandford-Znajek mechanism \citep{bz77} and might also exert a
torque on an accretion disc which happens to be tilted with respect to the BH
rotation axis \citep{bardeen75}.

The present work constitutes a follow-up to the analytic accretion model
introduced within a Newtonian framework by \cite{mendoza09} and for a
Schwarzschild spacetime by \cite{tejeda} (referred to in the following as
\hbox{Paper I}). Here we extend the general relativistic results of \hbox{Paper
I} to a Kerr spacetime. Our aim in this series of papers has been to construct
a toy model for the infall feeding an accretion disc around a BH, based solely
on the two leading ingredients determining the fluid bulk motion: gravity and
rotation. The model is based on the assumptions of stationarity, axisymmetry and
ballistic motion, i.e. we assume that the fluid particles follow geodesic lines
and neglect any deviation from their free-falling trajectories due to pressure
gradients, magnetic fields, self-gravity, radiative processes, etc. It is clear
that these assumptions constitute an oversimplification of the real situation
but they allow us to give a useful analytic description of the streamlines and
velocity field of the resulting flow. In addition, this will enable us to
highlight the signatures of pure relativistic effects in the accretion dynamics,
due to either the strong gravitational field or frame dragging, that might be
otherwise masked by a fully hydrodynamic treatment. Our analytic description of
the streamlines is based on the extensive body of work on geodesic motion
already existing in the literature \cite[see e.g,][and references
therein]{sharp,chandra}. Nevertheless, by using some standard identities for
Jacobi elliptic functions, we provide here a novel approach for writing the
solution for the radial and latitudinal motion of a timelike geodesic in Kerr
spacetime in terms of a single analytic expression.

There are several interesting astrophysical systems where the accretion regimes
are reasonably well approximated by the special conditions of the toy model.
For instance, \cite{beloborodov,kumar} and \cite{zalamea} used a ballistic
description for the infall feeding an accretion disc around a newly-formed BH,
following the collapse of a massive star in the so-called collapsar scenario
\citep{woosley}. \cite{kumar} described the infall within Newtonian theory while
\cite{beloborodov} and \cite{zalamea} made general relativistic studies, first
for a Schwarzschild BH and then for a Kerr one. These last two works were mainly
focused on numerically solving for the structure of the disc and predicting a
luminosity profile for the neutrino emission, while the infall was described
approximately, considering parabolic-like energies for the incoming particles
and restricting the analysis to boundary conditions with homogeneous density
and small, uniform rotation rates. On the other hand, full-hydrodynamic
numerical simulations for the collapsar scenario were performed by
\cite{lee06,camara09} and \cite{taylor11}. For some of the simulations presented
in those works, a fairly stationary regime is reached where the infalling matter
is not deviating significantly from free-fall in a large fraction of the spatial
domain of the simulation, but a shock front develops around the accretion disc
which abruptly decelerates the incoming streamlines and marks the transition
from an essentially ballistic regime to a hydrodynamical one \citep[see Paper I
for a comparison with one of the collapsar models in][]{lee06}. In those cases
the present analytic model might be a valuable tool for describing the infalling
matter which feeds the accretion disc, thereby enabling the exploration, in a
computationally efficient manner, of a wide range of (often uncertain) boundary
conditions (e.g. rotation profile, accretion rates) before performing full-scale
numerical simulations. 

In the present work we show comparisons between our analytic solutions and a
series of 3D smoothed particle hydrodynamics (SPH) simulations made using a
version of the publicly available code {\sc Gadget-2} \citep{springel}, modified by
\cite{taylor11} to include a simplified treatment of neutrino cooling and to
account for approximated general relativistic effects related to the Kerr metric
by using the second-order expansion pseudo-Newtonian potential developed by
\cite{MM}. The purpose of this comparison was twofold. Firstly, we wanted to
show the utility of the toy model as a simple, practical test for numerical
codes which include dynamical effects of general relativity. Simulation results
for pure ballistic motion (with the hydrodynamic forces being zeroed) can be
directly and quantitatively compared with exact general relativistic results
so as to test implementations of time-stepping algorithms, pseudo-Newtonian
potentials, etc. Secondly, once the numerical features of particle motion within
the simulation have been determined, one can then re-implement the hydrodynamic
features within the code to investigate the effects of fluid flow behaviour,
such as pressure, cooling and back-reaction from the accretion disc, on the
particle trajectories. This second point can be viewed as an exploration of the
validity of the ballistic approximation for describing the infall, showing the
utility of the toy model itself for understanding a wide variety of
astrophysical scenarios in a relatively simple way. 

The paper is organised as follows. The general setup of the model is described
in Section \ref{model}, while in Section \ref{velocity} the velocity field of
the accretion flow is given in terms of first integrals of motion. An analytic
description of the streamlines is given in Section \ref{lines} in terms of
Jacobi elliptic functions. A simple numerical scheme for calculating the density
field is presented in Section \ref{density}. Applications of the model for some
particular boundary conditions are then described in Section \ref{example} and
compared against full-hydrodynamic simulations in Section \ref{SPH}. Finally, a
general discussion and our conclusions are presented in Section
\ref{discussion}. Unless otherwise stated, we use geometrized units for which
$c=G=1$.

\setcounter{equation}{0}
\section{Model description}
\label{model}

In the present work we are constructing a model for the accretion flow of a
rotating cloud of non-interacting particles towards a Kerr BH of mass $M$ and
specific angular momentum $a$. The model is based on the assumptions of
stationarity and axisymmetry. We denote the constant accretion rate by
$\dot{M}$ (we are using the dot to represent differentiation with respect to 
proper time $\tau$). In order to describe the overall accretion flow we adopt 
the Boyer-Lindquist (BL) system of coordinates ($t,\ r,\ \theta,\ \phi$). The 
metric line element then has the familiar form \citep{mtw}
\begin{equation}
\begin{split}
 \ud s^2 = & -\left( 1 - \frac{ 2Mr }{ \rho^2 } \right)  \ud t^2 
- \frac{ 4\, aM r \sin^2 \theta }{ \rho^2 }\, \ud t\, \ud \phi \\ 
 & +\frac{ \rho^2 }{ \Delta }\, \ud r^2
+ \rho^2 \ud \theta^2 + \frac{ \Sigma \sin^2 \theta }{ \rho^2 }\, \ud \phi^2,
\end{split}
\label{e1.1}
\end{equation}
 where
\begin{equation}
\begin{split}
\rho^2 = r^2 &+ a^2\cos^2\theta,\qquad \Delta=r^2-2M r + a^2,\\
& \Sigma = \left(r^2 + a^2\right)^2 -a^2 \Delta \sin^2\theta.
\end{split}
\label{e1.2}
\end{equation}

Considering a set of non-interacting particles means, in practice, that we are
making a ballistic treatment of the fluid flow and hence that the accretion
dynamics are solely determined by the gravitational field of the BH. Under the
ballistic approximation, it is convenient to describe the whole gas cloud as a
collection of equal-mass test particles. We take as the boundary of our model a
`spherical'\footnote{A surface $r=$ const.\ in Kerr spacetime (in BL
coordinates) defines a spheroid with cross section $\left(\sqrt{r^2+a^2}
\sin\theta,\, r\cos\theta\right)$ in the $R$-$z$ plane (see Eqs.\,\ref{e8.5} and
\ref{e8.6}).} shell at $r=r_\s{0}$ from which test particles are continuously
injected. The infalling particles end up being either incorporated into an
infinitesimally thin equatorial disc or directly accreted inside the BH horizon
(located at $r_\s{+} = M +\sqrt{M^2-a^2}$). The analytic description of the
infall does not include the disc itself where clearly a ballistic treatment is
no longer valid; we shall just consider both disc and BH as passive sinks of
particles and energy. We take the particle properties at $r_\s{0}$ to be given
by specified distribution functions:
\begin{gather}
n_\s{0} = n\left(r_\s{0},\,\theta_\s{0}\right),\label{e1.3}\\
\dot{r}_\s{0} =
\dot{r}\left(r_\s{0},\,\theta_\s{0}\right),\label{e1.4}\\
\dot{\theta}_\s{0} =
\dot{\theta}\left(r_\s{0},\,\theta_\s{0}\right),\label{e1.5}\\
\dot{\phi}_\s{0}
=\dot{\phi}\left(r_\s{0},\,\theta_\s{0}\right), \label{e1.6}
\end{gather}
where $n$ is the particle number density (as measured in a co-moving
reference frame) and $\dot{r}$, $\dot{\theta}$, $\dot{\phi}$ are the radial,
polar and azimuthal components of the four-velocity, respectively.

We require the four distribution functions in Eqs.\,\eqref{e1.3}-\eqref{e1.6}
to be differentiable and symmetric with respect to the equatorial plane.
Additionally, in order to avoid streamlines intersecting before they reach the
equatorial plane, two further conditions need to be fulfilled. First, we require
that the test particles do not have turning points in their polar or radial
motion as they descend towards the equatorial plane. Second, we require the
mapping $\theta_\s{0} \rightarrow \theta$ to be non-singular, i.e.
\begin{equation}
\left.\left(\frac{\partial\theta}{\partial\theta_\s{0}}\right)\right|_r \ge 0.
\label{e1.7}
\end{equation}
A sufficient condition for satisfying Eq.\,\eqref{e1.7} for a large fraction of
the spatial domain is that the angular momentum distribution in the initial
shell should increase monotonically towards the equatorial plane. For initial
inward radial velocities above a certain threshold value, neighbouring
streamlines are found to intersect each other in the immediate vicinity of the
accretion disc. However the caustic surface defined by these intersections
remains close to the equatorial plane in a region where the ballistic
approximation is no longer expected to be valid \citep[the existence of this
detached caustic surface is also found in the Newtonian context as described
in][]{mendoza09}.

\setcounter{equation}{0}
\section{Velocity field}
\label{velocity}

Within the ballistic approximation, streamlines of the accretion flow correspond
to timelike geodesics in Kerr spacetime. Consider a test particle freely falling
towards the central BH with four-velocity $u^\s{\mu} = \dot{x}^\s{\mu}$. The
stationarity and axisymmetry of the Kerr metric lead to the existence of four
first integrals of the motion which will enable us to describe the velocity
field of the accretion flow \citep[for a detailed derivation of the constants of
motion, see][]{carter}. From the conservation of rest mass, one has as a first
conserved quantity the four-velocity modulus given as  
\begin{equation}
u^\s{\mu}\,u_\s{\mu} = -1.
\label{e2.1}
\end{equation}
The other three constants of motion are: $E$, the total specific energy; $\ell$, 
the projection of the specific angular momentum on to the BH rotation axis; and 
$Q$, the Carter constant. Using BL coordinates, these quantities are given by
\begin{gather}
E  = \left( 1 - \frac{ 2M r }{ \rho^2 } \right)  \dot{t} + 
\frac{ 2\,aM r \sin^2 \theta }{ \rho^2 } \dot{\phi} ,
\label{e2.2} \\
\ell  = - \frac{ 2\,aM r \sin^2 \theta }{ \rho^2 }\, \dot{t} +
\frac{ \Sigma \sin^2 \theta }{ \rho^2 }\, \dot{\phi},
\label{e2.3} \\
Q  = \rho^4\dot{\theta}^2 + \ell^2 \cot^2\theta -\varepsilon\,a^2\cos^2\theta,
\label{e2.4}
\end{gather}

\noindent where, for convenience, we have introduced 
\begin{equation}
\varepsilon  = E^2-1. 
\label{e2.5}
\end{equation}

Once the boundary conditions in Eqs.\,\eqref{e1.3}-\eqref{e1.6} have been fixed,
the conserved quantities in Eqs.\,\eqref{e2.2}-\eqref{e2.4} are completely
determined. Nonetheless, note that in general the conserved quantities will be
functions of the initial polar angle $\theta_\s{0}$ and, hence, vary from
streamline to streamline. The value of $\dot{t}_\s{0}$ is calculated from
Eq.\,\eqref{e2.1} as a function of $\dot{r}_\s{0}$, $\dot{\theta}_\s{0}$ and
$\dot{\phi}_\s{0}$. Making use of the four integrals of motion, one gets the
following system of equations determining the proper time evolution of the
coordinates of the particle:
\begin{gather}
\rho^2 \frac{\ud r }{\ud\tau} =  \pm \sqrt{\mathcal{R}},
\label{e2.6}\\
\rho^2 \frac{\ud \theta }{\ud\tau} =  \pm \sqrt{\Theta},
\label{e2.7}\\
\rho^2 \frac{\ud \phi }{\ud\tau} =  \frac{\mathcal{A}}{\sin^2\theta} +
a\,\mathcal{B},
\label{e2.8}\\
\rho^2 \frac{\ud t }{\ud\tau} =  a\,\mathcal{A} +
\mathcal{B}\left(r^2+a^2\right),
\label{e2.9}
\end{gather}
where
\begin{gather}
\mathcal{R} = \varepsilon\, r^4 + 2Mr^3 + \left( \varepsilon\,a^2 - \ell^2 - 
Q\right) r^2 \nonumber\\
\hspace{1.4cm} + 2M\left[ Q + \left( a\,E - \ell \right)^2\right]r-a^2 Q,
\label{e2.10}\\
\Theta =  Q +\varepsilon\,a^2\cos^2\theta-\ell^2\cot^2\theta, 
\label{e2.11}\\
\mathcal{A} = \ell-a\,E\,\sin^2\theta,
\label{e2.12}\\
\mathcal{B} =  \left[ E\left(r^2+a^2\right) - a\,\ell \right]/ \Delta .
\label{e2.13}
\end{gather}

\noindent The signs in Eqs.\,\eqref{e2.6} and \eqref{e2.7} are independent of
each other and change whenever the test particle reaches a turning point in its
trajectory (though, recall that in the present scenario, the radial coordinate
has been required to decrease monotonically as the particle approaches the
equatorial plane). Regarding the polar motion, since we have assumed mirror
symmetry with respect to the equatorial plane, we can consider without loss of
generality that the particle on which we are focusing is in, say, the northern
hemisphere, i.e. $0<\theta_\s{0}<\pi/2$. For such a particle the polar
coordinate increases from $\theta_\s{0}$ to $\pi/2$. We then take the minus sign
in Eq.\,\eqref{e2.6} and the plus sign in Eq.\,\eqref{e2.7}.

In Eqs.\,\eqref{e2.6}-\eqref{e2.9}, we already have expressions for the four
components of the velocity field:
\begin{gather}
u^\s{r} =  - \frac{\sqrt{\mathcal{R}}}{\rho^2},
\label{e2.14}\\
u^\s{\theta} =  \frac{\sqrt{\Theta}}{\rho^2},
\label{e2.15}\\
u^\s{\phi} = \frac{\mathcal{A} +
a\,\mathcal{B}\,\sin^2\theta}{\rho^2\sin^2\theta},
\label{e2.16}\\
u^\s{t} =  \frac{a\,\mathcal{A} +
\mathcal{B}\left(r^2+a^2\right)}{\rho^2}.
\label{e2.17}
\end{gather}

Eqs.\,\eqref{e2.14}-\eqref{e2.17} represent velocities with respect to the 
BL coordinate system as opposed to physical, locally measured ones. In order to
get a local description of the velocity field, we follow \cite{bardeen72} and
introduce a set of locally non-rotating frames (LNRFs). Associated with each
LNRF there is an orthonormal tetrad of four-vectors constituting a local
Minkowskian coordinate set of basis vectors. If we denote with a bar coordinates
with respect to this local reference frame, the corresponding physical
three-velocity field is given by
\begin{gather}
 V^\s{\bar{r}} =  - \frac{\sqrt{\mathcal{R}/\Delta}}{\gamma\,\rho},
\label{e2.18}\\
 V^\s{\bar{\theta}} = \frac{\sqrt{\Theta}}{\gamma\,\rho}, 
\label{e2.19}\\
 V^\s{\bar{\phi}} = \frac{\rho\,\ell}{\gamma\,\sqrt{\Sigma}\,\sin\theta},
\label{e2.20}
\end{gather}
where $\gamma$ is the Lorentz factor between the LNRF and the test
particle passing by, and is given by 
\begin{equation}
\gamma = \sqrt{1+\frac{\mathcal{R}}{\Delta\,\rho^2}+
\frac{\Theta}{\rho^2}+\frac{\rho^2\,\ell^2}{\Sigma\sin^2\theta}}.
\label{e2.21}
\end{equation}

Note that both sets of expressions for the velocity field (Eqs.\,\ref{e2.14} -
\ref{e2.17} and Eqs.\,\ref{e2.18} - \ref{e2.20}) are functions of the position
($r$, $\theta$) as well as of the conserved quantities along each streamline,
which have been determined by the initial position ($r_\s{0}$, $\theta_\s{0}$).
Therefore, to use them in practice we need to provide an explicit mapping from
($r_\s{0}$, $\theta_\s{0}$) $\mapsto$ ($r$, $\theta$). Such a mapping will be
given in the next section in terms of an expression for the streamlines.

\setcounter{equation}{0}
\section{Streamlines}
\label{lines}

In this section we give an analytic solution for the radial and latitudinal 
motion of test particles freely falling in Kerr spacetime. It is not within the
scope of the present work to give an exhaustive discussion of the qualitative
features of the trajectories. For a thorough discussion of timelike geodesics
around a rotating BH, see \cite{wilkins,bardeen73,chandra, dymnikova,novikov,
kraniotis,fujita,levin12}, while an in-depth analysis of the latitudinal and
radial motion is given by \cite{felice72, bicak76}. In the following, we build
on these previous works and give a novel approach for expressing the radial and
latitudinal solutions of timelike geodesic motion by means of a single
analytical formula. 

Given the assumptions of stationarity and axisymmetry, all that is needed for
completely describing a streamline of the present model is to consider the
projection of an arbitrary timelike geodesic onto the $r$-$\theta$ plane. For
doing this, it is sufficient to consider Eqs.\,\eqref{e2.6} and \eqref{e2.7} and
to combine them in the following way
\begin{equation}
\mathlarger{\int}^{ r }_{ r_\s{0} } \frac{ \ud r' }{\sqrt{\mathcal{R}(r')}}=-
\mathlarger{\int}^{ \theta }_{ \theta_\s{0} } \frac{ \ud \theta'}
{\sqrt{\Theta(\theta')}}.
\label{e3.1}
\end{equation}

\noindent The solutions to both sides of the above equation can be expressed in
terms of elliptic integrals \citep[see e.g.][]{byrd}. Let us introduce the
following definitions:
\begin{gather}
\Phi(r) = \mathlarger{\int}^{ r }_{ r_\s{a} } 
\frac{ \ud r'}{\sqrt{\mathcal{R}(r')}}, \label{e3.2} \\
\Psi(\theta) = \mathlarger{\int}^{ \theta }_{ \theta_\s{a} } 
\frac{ \ud \theta'}{\sqrt{\Theta(\theta')}}, \label{e3.3}
\end{gather}

\noindent where $r_\s{a}$ and $\theta_\s{a}$ are as yet unspecified reference
points in the particle trajectory. With these definitions we can rewrite
Eq.\,\eqref{e3.1} as
\begin{equation}
\Phi(r) - \Phi(r_\s{0}) = \Psi(\theta_\s{0}) - \Psi(\theta).\label{e3.4}
\end{equation}
 
\noindent In the following, we give explicit expressions for $\Phi(r)$ and
$\Psi(\theta)$.

\subsection{Radial solution}
\label{radial}

Consider first Eq.\,\eqref{e3.2}. The procedure for solving this integral is
technically the same as in the Schwarzschild case (see Paper I), with the
solution depending on the nature of the roots of $\mathcal{R}(r)$. The physical
interpretation of these roots is clear: whenever they are real and greater than
$r_\s{+}$, they constitute turning points of the radial motion at which
$\dot{r}$ changes sign and the direction of integration for the radial integral
reverses. In the following, we briefly review the results of \cite{wilkins} and
\cite{dymnikova} concerning the properties of the roots of $\mathcal{R}(r)$.

Since $\mathcal{R}(r)$ is a fourth order polynomial, there are the following
possibilities for the roots: all four are real; two are real while the other two
form a complex conjugate pair; there are two pairs of complex conjugates. The
first two cases include the possibility of multiplicity of the real roots.
Although one can express the roots analytically in terms of the parameters of
the orbit ($E$, $\ell$ and $Q$) \citep[see e.g.][]{abramowitz}, we do not give
the final expressions coming from this here because we do not find them
particularly useful in practice. Instead, we assume that we have already found
the four roots, either analytically or by means of a root finding algorithm, and
write $\mathcal{R}(r)$ as 
\begin{equation}
\mathcal{R}(r) = \varepsilon(r-r_\s{a})(r-r_\s{b})(r-r_\s{c})(r-r_\s{d}).
\label{e4.1}
\end{equation}

\noindent We label the roots in the following way:

\vspace{12pt}
{\bf (i)} If $\mathcal{R}(r)$ has four real roots, and in order to satisfy the
condition $\mathcal{R}(r)>0$, there are two possibilities: either $r$ is
bracketed in between two non-negative consecutive roots of $\mathcal{R}(r)$, or
$r$ has a lower bound and is unbounded above. The latter case represents an open
orbit with the largest positive root being the only turning point. 

In the first case, we call the roots bracketing $r$, $r_\s{a}$ and $r_\s{b}$
(with $r_\s{a}<r_\s{b}$). In the second case we again take $r_\s{a}$ as the
lower bound for $r$ and let $r_\s{b}$ be the negative root with the largest
absolute value. In both cases the two remaining roots are denoted as $r_\s{c}$
and $r_\s{d}$ (with $|r_\s{c}| < |r_\s{d}|$). 

\vspace{12pt}
{\bf (ii)} If $\mathcal{R}(r)$ has two real roots and a complex conjugate pair,
we take $r_\s{a}$ and $r_\s{d}$ (with $|r_\s{a}|<|r_\s{d}|$) to be the real
roots while $r_\s{b}$ and $r_\s{c}$ form the complex conjugate pair. 

\vspace{12pt}
{\bf (iii)} If $\mathcal{R}(r)$ has two complex conjugate pairs of roots, we
take $r_\s{b} = r_\s{c}^*$ and $r_\s{a}=r_\s{d}^*$ with $\text{Re}(r_\s{a}) <
\text{Re}(r_\s{b})$. Note that this possibility is a special characteristic of
Kerr spacetime, since in the Schwarzschild case one of the roots was zero and
hence at least one other root was real as well.

\vspace{12pt}
With this way of labelling the roots, we can now express the radial solution as
\citep{byrd} 
\begin{gather}
  \Phi(r) = \frac{2\,\cn^{-1}\left(\sqrt{\frac{(r_\s{d}
-r_\s{a})(r_\s{b}-r)}{(r_\s{b}-r_\s{a})(r_\s{d}-r)}},\,k_\s{r}\right)}
{\sqrt{\varepsilon(r_\s{a}-r_\s{c})(r_\s{d}-r_\s{b})}},\label{e4.2}\\
  k_\s{r} = \sqrt{\frac{(r_\s{b}-r_\s{a})(r_\s{d}-r_\s{c})}
{(r_\s{d}-r_\s{b})(r_\s{c}-r_\s{a})}}.
\label{e4.3}
\end{gather}

\noindent where $\cn^{-1}(u,\,k_\s{r})$ is the inverse of the Jacobi elliptic
function $\cn(\varphi,\,k_\s{r})$ with modulus $k_\s{r}$. We refer interested
readers to specialised literature on elliptic integrals
\citep[e.g.][]{hancock,cayley,lawden} for a precise definition of Jacobi
elliptic functions in terms of elliptic integrals.

In the cases (ii) and (iii), where complex roots arise, intermediate steps in
the calculation of Eq.\,\eqref{e4.2} involve the use of complex quantities,
although the final result $\Phi(r) - \Phi(r_\s{0})$ is always a real number. For
alternative forms of the solution for the second and third cases, involving only
explicitly real terms, see the Appendix.

Finally, we note that the solution for the radial motion in Eq.\,\eqref{e4.2} is
formally identical to the solution given in Paper I for a Schwarzschild
spacetime.

\subsection{Polar angle solution}
\label{polar}

We next consider Eq.\,\eqref{e3.3}. From that equation it is clear that the
latitudinal motion is restricted to those values of $\theta$ such that
$\Theta(\theta)\ge 0$. Just as in the radial case, the polar angle solution
depends on the nature of the roots of the equation $\Theta(\theta) = 0$.
Whenever these roots belong to the natural domain of $\theta$, i.e.
$\theta\in[0,\pi]$, they constitute  turning points of the polar motion at which
$\dot{\theta}$ changes sign. These roots are straightforward to obtain after
noticing that the equation $\Theta(\theta) = 0$ is equivalent to the following
quadratic polynomial equation in $\cos^2\theta$ 
\begin{equation}
 \varepsilon\,a^2\cos^4\theta+(Q+\ell^2-\varepsilon\,a^2)\cos^2\theta-Q =0.
\label{e5.1}
\end{equation}

Let $\theta_\s{a}$ be the turning point of the polar motion closest to the polar
axis. For $\ell\ne0$, $\theta_\s{a}$ corresponds to the smallest positive root
of Eq.\,\eqref{e5.1}, and in this case it is convenient to rewrite $Q$ in terms
of $\theta_\s{a}$ as
\begin{equation}
Q = \ell^2\cot^2\theta_\s{a} - \varepsilon\,a^2\cos^2\theta_\s{a}.
\label{e5.2}
\end{equation}

In general, the polar equation $\Theta(\theta) = 0$ will have zero, two or four
real roots in the interval of interest. The first case arises when $\ell=0$ and
$Q>-\varepsilon a^2$, corresponding to a test particle sweeping the whole polar
domain and periodically crossing the polar axis.\footnote{Even though in this
case the angles $\theta=0,\,\pi$ do not satisfy $\Theta(\theta) = 0$, they still
represent turning points since at those locations the polar velocity changes
sign discontinuously. When this happens, we take $\theta_\s{a}=0$.} The second
case corresponds to bounded polar motion, $\theta\in[\theta_\s{a}, \pi -
\theta_\s{a}]$, with the test particle periodically crossing the equatorial
plane. The third case corresponds to a test particle restricted to move within a
single hemisphere (in this case the northern one) as $\theta\in[ \theta_\s{a},
\theta_\s{b}]$, where $\theta_\s{b}\le\pi/2$ is the second turning point of the
polar motion.
 
As follows from Eqs.\,\eqref{e2.6} and \eqref{e2.7}, the polar and radial
motions are decoupled, and hence their turning points are in general independent
of each other. Since we have assumed that the particle starts its journey from
the northern hemisphere, we have $\theta_\s{a}\le \theta_\s{0} \le\pi/2$. 

In terms of the angle $\theta_\s{a}$, the solution for the polar integral is
given by \citep{byrd} 
\begin{gather}
\Psi(\theta) = \frac{\cos\theta_\s{a}}{\sqrt{Q}}\,\cd^{-1}
\left(\frac{\cos\theta}{ \cos\theta_\s{a}},\, k_\s{\theta}\right),
\label{e5.3} \\
k_\s{\theta} = \sqrt{- \varepsilon\,a^2/Q}\, \cos^2\theta_\s{a},
\label{e5.4}
\end{gather}

\noindent where $\cd^{-1}(u,\,k_\s{\theta})$ is the inverse of the Jacobi
elliptic function $\cd(\varphi,\,k_\s{\theta})$ with modulus $k_\s{\theta}$. 

Note that $\Psi(\theta_\s{a})=0$, while for $\theta=\pi/2$
\begin{equation}
\Psi(\pi/2)=\frac{\cos\theta_\s{a}}{\sqrt{Q}}\,K(k_\s{\theta}) ,
\label{e5.5}
\end{equation}

\noindent where $K(k_\s{\theta})$ is the complete elliptic integral of the first
kind. 

Just as in the radial case, for some values of the parameters $\varepsilon$ and
$Q$, intermediate steps in the computation of Eq.\,\eqref{e5.3} might involve
the use of complex quantities but the final result is always a real quantity.
See the Appendix for alternative expressions for the polar solution involving
just real quantities. 

For the non-rotating BH case ($a=0$) one gets $k_\s{\theta}=0$ from
Eq.\,\eqref{e5.4}. On the other hand, for a null value of the modulus, one has 
that $\cd(\varphi,0) =\cos(\varphi)$, and therefore in this case
Eq.\,\eqref{e5.3} can be simplified as
\begin{equation}
\Psi(\theta) = \frac{\cos\theta_\s{a}}{\sqrt{Q}}\,\cos^{-1}
\left(\frac{\cos\theta}{ \cos\theta_\s{a}}\right),
\label{e5.6}
\end{equation} 

\noindent which is the same expression as that found in Paper I (following a
different approach).

\subsection{Timelike geodesics}

Bringing together the results in Eqs.\,\eqref{e4.3} and \eqref{e5.3}, we arrive
at the following expression for the projection onto the $r$-$\theta$ plane of a
timelike geodesic in Kerr spacetime:
\begin{equation}
r = \frac{r_\s{b}(r_\s{d}-r_\s{a})-r_\s{d}(r_\s{b}-r_\s{a}) 
\cn^2\left(\xi,\,k_\s{r}\right) }
{r_\s{d}-r_\s{a}-(r_\s{b}-r_\s{a})\cn^2\left(\xi,\,k_\s{r}\right) },
\label{e6.1}
\end{equation}

\noindent with
\begin{equation}
\xi = \frac{ \sqrt{\varepsilon(r_\s{a}-r_\s{c})(r_\s{d}-r_\s{b}) } }{ 2 } 
\left[\Phi(r_\s{0})+\Psi(\theta_\s{0})-\Psi(\theta)\right].
\label{e6.2}
\end{equation}

\noindent Eqs.\,\eqref{e6.1} and \eqref{e6.2} constitute the analytic expression
for the streamline of a gas element being accreted from ($r_\s{0}$,
$\theta_\s{0}$).

\setcounter{equation}{0}
\section{Density field}
\label{density}

In order to get a formal expression for calculating the density field, we start
from the continuity equation 
\begin{equation}
 \left( n\,u^\s{\mu} \right)_\s{;\mu}  = 0,
\label{e7.1}
\end{equation}

\noindent where a semi-colon denotes covariant differentiation. We integrate
the above expression over a four-volume element $\mathcal{V}$, consisting of a
streamline tube extending for an infinitesimal interval of coordinate time $\ud
t$. We take the spatial cross-section of this streamline tube to be the
collection of all of the streamlines starting to fall-in from a differential
area element $\ud x^2|_\s{r_\s{0}}$ at the initial shell and ending up at a
second sphere with arbitrary radius $r<r_\s{0}$. Denoting by
$\partial\mathcal{V}$ the hypersurface delimiting the integrating volume and
invoking the Gauss theorem, we have that
\begin{equation}
  \underset{\mathcal{V}}{\int} \left( n\,u^\s{\mu} \right)_\s{;\mu}
\sqrt{-g}\,\ud^4 x = 
  \underset{\partial\mathcal{V}}{\oint} n\,u^\s{\mu} N_\s{\mu}
\sqrt{|h|}\ud^3 x = 0,
\label{e7.2}
\end{equation}

\noindent where $N_\s{\mu}$ is a unit vector normal to $\partial\mathcal{V}$ and
$h$ is the determinant of the induced metric on this hypersurface. Since we have
assumed stationarity, it is clear that the net particle flux through any closed
spatial hypersuface at a given time $t$ equals zero. Moreover, for the remaining
mixed time-space hyperfaces of $\partial\mathcal{V}$, the contraction
$u^\s{\mu}N_\s{\mu}$ will be, by construction, different from zero just for a
hypersurface oriented perpendicularly to the radial direction. Hence, we have
that Eq.\,\eqref{e7.2} reduces to  
\begin{equation}
  \left.n\,u^\s{\mu} N^{(r)}_\s{\mu}
\sqrt{|h^{(r)}|}\ud t\,\ud\theta\,\ud\phi\right|_r^{r_\s{0}} = 0.
\label{e7.3}
\end{equation}

\noindent Substituting into this equation that $N^{(r)}_\s{\mu}=
\delta^r_\s{\mu}/ \sqrt{g^\s{rr}}$ where $g^\s{rr}=\Delta/\rho^2$, together with
$h^{(r)} = \Delta\rho^2\sin^2\theta$, we arrive at
\begin{equation}
\left.n\,u^\s{r}\, \rho^2\,\sin \theta\,
\ud t\,\ud\theta\,\ud\phi\right|_r^{r_\s{0}} = 0.
\label{e7.4}
\end{equation}

\noindent Invoking once again the stationarity and axisymmetry conditions, it
follows that $\ud t_\s{0}\,\ud\phi_\s{0} = \ud t\,\ud\phi$. Using this result in
Eq.\,\eqref{e7.4} allows us to solve for $n$, getting 
\begin{equation}
 n = \frac{n_\s{0}\,u^r_\s{0}\,\rho^2_\s{0}\,\sin \theta_\s{0}}
{u^r\, \rho^2\,\sin \theta }
\left(\frac{\partial \theta}{\partial \theta_\s{0}}\right)^{-1}.
\label{e7.5}
\end{equation}

Just as in the Schwarzschild case, analytically calculating the partial
derivative in Eq.\,\eqref{e7.5} would be a rather involved process, whereas 
evaluating it numerically is a trivial task. We refer the reader to Paper I for
a description of a numerical scheme for computing $n$.

The requirement that there should be no early intersections of streamlines
before the equator has been reached (see Eq.\,\ref{e1.7}) ensures that the
expression for calculating the density in Eq.\,\eqref{e7.5} is well defined.

\setcounter{equation}{0}
\section{Applications of the analytic model}
\label{example}

We now illustrate our analytic model by applying it to an example scenario with
boundary conditions consisting of matter in uniform rotation on a uniform shell,
i.e.:
\begin{align}
& n_\s{0} = \text{const.},\label{e8.1}\\
& \dot{r}_\s{0} = \text{const.},\label{e8.2}\\
& \dot{\phi}_\s{0} =\text{const.},\label{e8.3}\\
& \dot{\theta}_\s{0} = 0. \label{e8.4}
\end{align}

\noindent The condition in Eq.\,\eqref{e8.4} implies that, for every streamline,
$\theta_\s{a} = \theta_\s{0}$.

\begin{figure*}
\begin{center}
  \includegraphics[height=215mm]{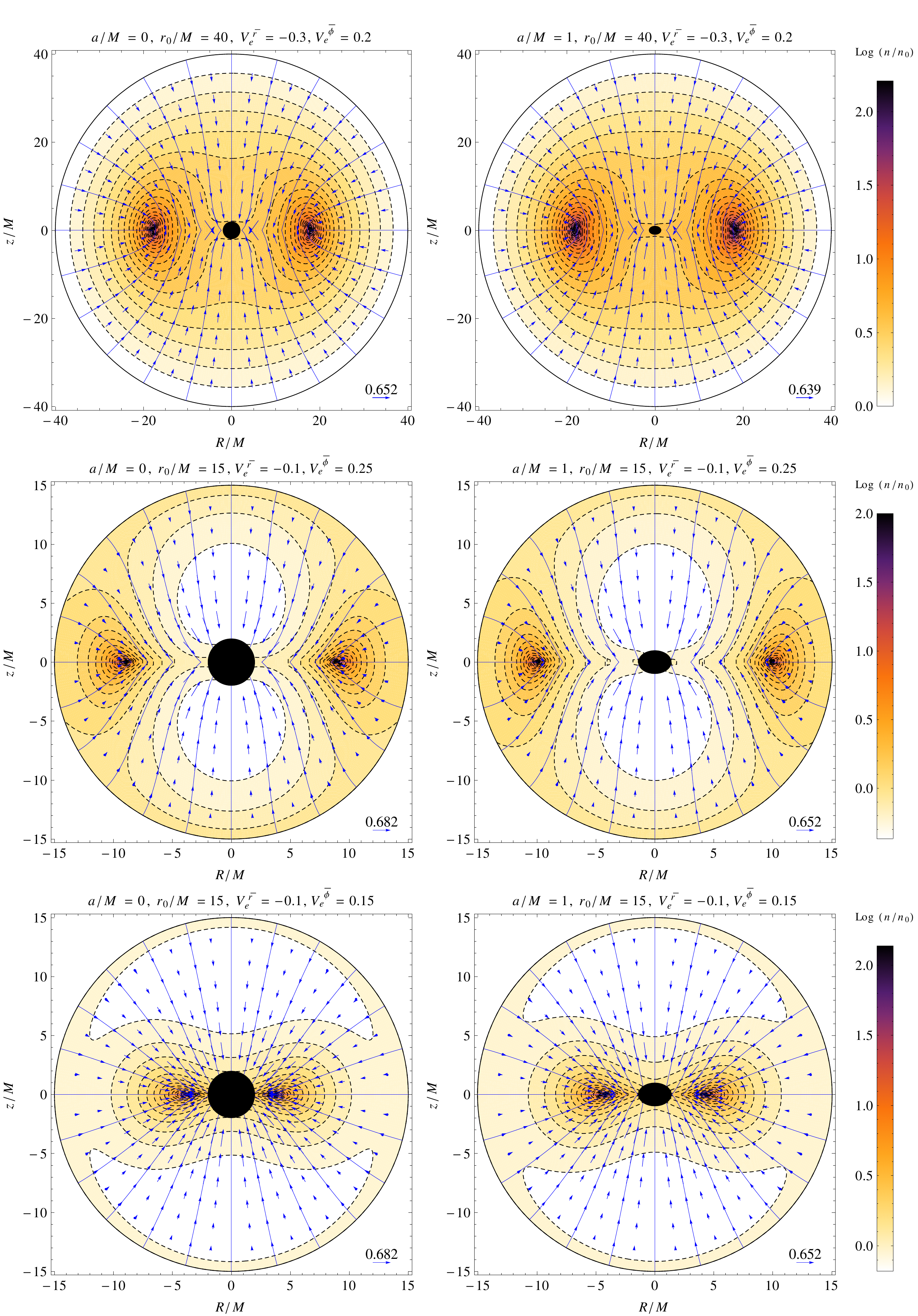}
\end{center}
 \caption{Streamlines, velocity field and density contours for six different
combinations of the flow parameters. The values of the parameters used in each
case are indicated above each panel. Each panel shows the spatial projection
onto the $R$-$z$ plane and the colour coding corresponds to the value
of the logarithm of the particle number density Log$(n/n_\s{0})$, with the scale
being indicated by the colour-coding bar at the left of each row. The arrows
correspond to the $V^\s{\bar{r}}$ and $V^{\bar{\theta}}$ components of the
velocity field. The magnitude of the largest arrow is indicated at the bottom
right of each panel.}
 \label{f1}
\end{figure*}

Figure~\ref{f1} shows six panels with the streamlines, velocity field and
density contours for six different combinations of the flow parameters. The
panels consist of spatial projections onto the $R$-$z$ plane, where $R$ and $z$
(together with $\theta$) are the cylindrical coordinates associated with
the BL ones and are defined as 
\begin{align}
 R & = \sqrt{r^2+a^2}\,\sin\theta, \label{e8.5}\\
 z & = r\,\cos\theta. \label{e8.6}
\end{align}

\noindent For specifying the set of model parameters in each case, we have used
$a$, $r_\s{0}$, $V_e^\s{\bar{r}}$ and $V_e^\s{\bar{\phi}}$, where the subscript
$e$ indicates that the corresponding quantity is being evaluated at the equator
of the shell. Note that, for the present boundary conditions, the set of
parameters ($a$, $r_\s{0}$, $V_e^\s{\bar{r}}$, $V_e^\s{\bar{\phi}}$) has a
one-to-one correspondence with ($a$, $r_\s{0}$, $\dot{r}_\s{0}$,
$\dot{\phi}_\s{0}$) given by the inversion of the system of
Eqs.\,\eqref{e2.18}-\eqref{e2.21}. Also note that fixing this set of parameters
specifies a family of models rather than a single one, since both length and
density scales can still be arbitrarily and independently chosen.

\begin{figure}
\begin{center}
  \includegraphics[width=84mm]{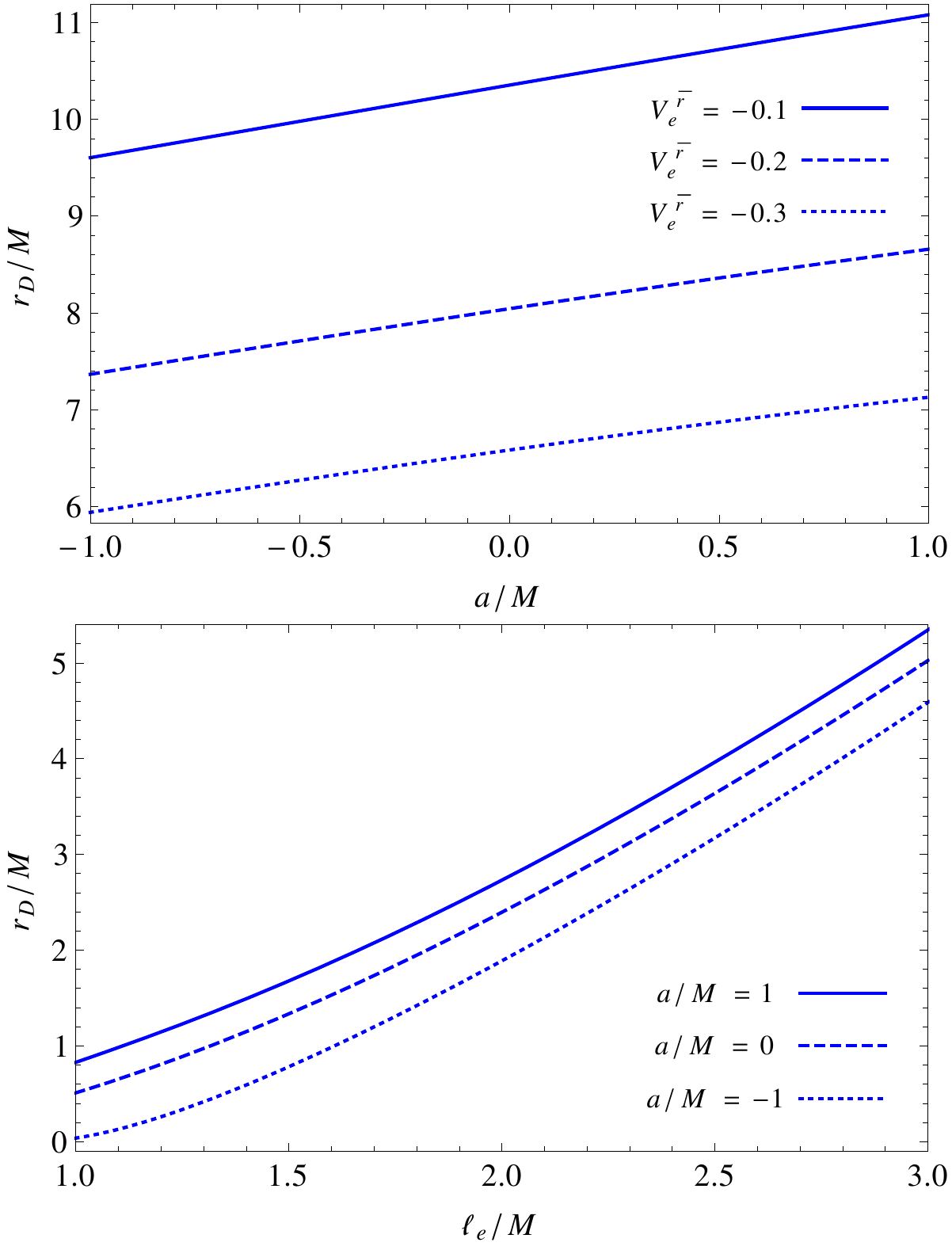}
\end{center}
 \caption{Plot of $r_\s{D}$ versus $a$ (top panel) and $r_\s{D}$ versus $\ell_e$
(bottom panel). In the top panel, fixed values are taken for $r_\s{0}\ (=20\,M)$
and $V_e^\s{\bar{\phi}}\ (=0.2)$ while in the bottom panel, fixed values are
taken for $r_\s{0}\ (=60\,M)$ and $V_e^\s{\bar{r}}\ (= -0.2)$. Note how the BH
spin couples with the angular momentum of the infalling matter and leads to a
larger $r_\s{D}$ for a co-rotating flow and to a smaller $r_\s{D}$ for a
counter-rotating one.}
 \label{f2}
\end{figure}

The radius of the outer edge of the disc being formed, $r_\s{D}$, measured as 
the outermost material first reaches the equatorial plane, can be calculated 
from Eq.\,\eqref{e6.2}, taking first $\theta=\pi/2$ and then 
$\theta_\s{0}=\pi/2$, giving
\begin{equation}
\xi_\s{D} = \frac{\sqrt{\varepsilon(r_{\s{a}}-r_{\s{c}})
(r_{\s{d}}-r_{\s{b}}) }}{2}\left[ \Phi(r_\s{0}) -
\frac{\pi}{2\sqrt{\ell_e^2-\varepsilon\,a^2}} \right] ,
\label{e8.7}
\end{equation}

\noindent and then substituting the result into Eq.\,\eqref{e6.1}. 

In Figure~\ref{f2}, we have plotted $r_\s{D}$, first as a function of the BH
spin $a$ and then as a function of the specific angular momentum at the equator
of the shell, $\ell_e=\ell(\pi/2)$. Here we have assumed $V_e^\s{\bar{\phi}}>0$,
and so a negative value of $a$ implies a counter-rotating disc. From this figure
we can clearly see how the BH spin couples with the angular momentum of the disc
(through the frame dragging effect), giving rise to a larger $r_\s{D}$ for a
co-rotating disc and a smaller $r_\s{D}$ for a counter-rotating one. It is also
clear that, as intuitively expected, $r_\s{D}$ is a monotonically increasing
function of both $V_e^\s{\bar{\phi}}$ and $V_e^\s{\bar{r}}$.

\begin{figure}
\begin{center}
  \includegraphics[width=84mm]{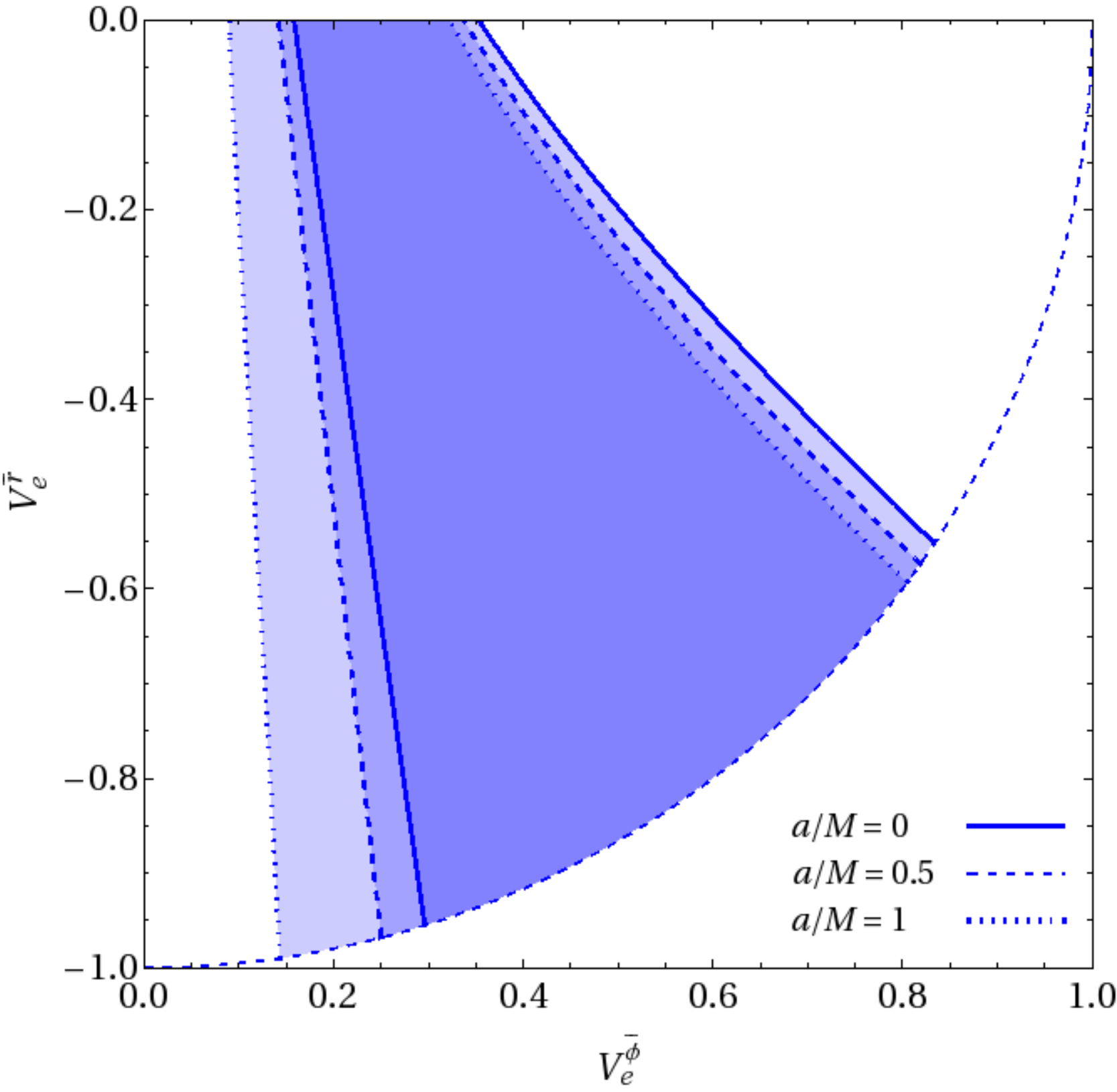}
\end{center}
 \caption{The pairs of velocity values $V_e^\s{\bar{r}}$-$V_e^\s{\bar{\phi}}$
leading to a disc radius such that $r_\s{D}\in[r_\s{+},r_\s{0}]$ are plotted for
a fixed value of $r_\s{0}=10\,M$ and for $a/M=0,\,0.5,\,1$. The upper boundary
for each value of $a$ represents the points ($V_e^\s{\bar{r}},
V_e^\s{\bar{\phi}}$) such that $r_\s{D}=r_\s{0}$, while the lower one represents
those such that $r_\s{D}=r_\s{+}$. Note how the domain of values in the velocity
space leading to physically relevant accretion models shifts to smaller values
of $V_e^\s{\bar{\phi}}$ as $a$ increases, because of the frame-dragging effect.}
\label{f3}
\end{figure}

Working with the LNRF velocities $V_e^\s{\bar{r}}$ and $V_e^\s{\bar{\phi}}$
makes the exploration of the parameter space easier since, being physical
velocities, they are naturally bounded as $V_e^\s{\bar{r}}\in(-1,0]$ and
$V_e^\s{\bar{\phi}}\in[0,1)$. Furthermore, for fixed values of $r_\s{0}$ and
$a$, a pair of velocities in the $V_e^\s{\bar{r}}$-$V_e^\s{\bar{\phi}}$ plane is
also restricted by the condition that the resulting $r_\s{D}$ should satisfy
$r_\s{D}\in(r_\s{+},r_\s{0})$. In Figure~\ref{f3} we have plotted the regions
on the velocity space which lead to an outer radius of the disc satisfying this
criterion. The plot has been constructed for a fixed value of $r_\s{0}=10\,M$
and three different values of $a$. From this figure we observe that the domain
of values in the velocity space leading to physically relevant accretion models
shifts to smaller values of $V_e^\s{\bar{\phi}}$ as $a$ increases. This
behaviour is a consequence of the frame dragging effect: for a given test
particle with fixed azimuthal velocity $V_e^\s{\bar{\phi}}$, its associated
angular momentum is an increasing function of $a$, and hence points in the
$V_e^\s{\bar{r}}$-$V_e^\s{\bar{\phi}}$ plane which, in the low-$a$ case,
did not have large enough angular momentum to keep the outer edge of the disc
outside the event horizon, are able to do so for a larger value of $a$. 
Conversely, low-$a$ models with an angular momentum only just small enough to
form any disc inside their initial shell would have discs entirely outside their
initial shell when $a$ is increased (thus becoming excluded from the parameter
domain). Also note that this parameter-space effect is greater on the lower
boundary of $V_e^\s{\bar{\phi}}$ than on the upper one, which is simply due to
the fact that the frame dragging increases as $r\rightarrow r_\s{+}$.

\setcounter{equation}{0}
\section{Comparison with numerical simulations}
\label{SPH}

In this section we compare the analytic solution derived above against SPH
simulations performed with the modified version of the code {\sc Gadget-2}
\citep{springel} presented and used by \cite{taylor11}. In that work, the
authors studied numerically the production of the progenitors for long-duration
GRBs as the aftermath of the collapse of a massive star, starting off from
realistic initial conditions and including cooling by neutrino emission. They
also included a rough approximation of Kerr metric general relativistic effects
by using the second-order expansion pseudo-Newtonian potential developed by
\cite{MM} (MM). Note that, among the several different pseudo-Newtonian
potentials that exist for mimicking effects of Kerr spacetime, the choice of MM
was intended to roughly minimise the overall effect of errors in the
approximation of various dynamic properties such as location of the innermost
stable circular orbit, epicyclic frequencies and radial acceleration \citep[for
further discussion see the Appendix B of][]{taylor11}. 

Here, we compare our ballistic toy model (calculated with the full Kerr metric)
with results from various versions of the SPH code implementing approximate
gravity prescriptions: in addition to the MM potential, we have also
experimented with the classical Newtonian potential and the widely-used
pseudo-Newtonian potential of \cite{PW} (PW). In connection with the two
pseudo-Newtonian potentials, we should stress that they have been designed
particularly for capturing relevant relativistic features of importance for
accretion discs, including getting correct locations for the innermost stable
circular orbit in Kerr and Schwarzschild spacetimes, respectively. This does not
at all guarantee that they would be good for other purposes such as the infall
calculations being discussed here. However, they have been widely used in more
general contexts and so it is relevant for us to test them against the toy
model.

In order to perform a systematic analysis in which we are able to distinguish
between hydrodynamic and gravitational effects, we have considered two kinds of
simulation:

\vspace{12pt}
{\bf (i)} Ballistic free-fall, with the SPH particles being automatically
removed when they cross either the equatorial plane or the BH horizon. Here we
consider an equation of state (EoS) for which the fluid pressure $P=0$, and
hence, effectively `turn off' the hydrodynamical forces. The aim of this kind of
simulation is to highlight the differences in the flow dynamics coming from the
use of different gravity descriptions (full general relativity, Newtonian
gravity, and the MM and PW pseudo-Newtonian potentials) and from different
numerical implementations of the equations of motion for the particles. 
 
\vspace{12pt}
{\bf (ii)} Full-hydrodynamical simulations, including back reaction from a
growing equatorial disc and cooling in regions where the gas gets very hot
($\ge10^9$K). Here we do not remove SPH particles when they reach the equatorial
plane but rather let them settle down by themselves into a disc structure. In
this case we employ a polytropic EoS of the form $P=(\Gamma -1)\,n\,u$, where
$n$ is the baryon number density, $u$ the internal energy per baryon and
$\Gamma$ is the polytropic index. We take $\Gamma=4/3$, and the value of the
internal energy (taken to be constant in the initial shell) is set at an
arbitrary but non-negligible value of one tenth of the sum of kinetic energy and
absolute value of the Newtonian potential energy for an SPH particle at the
equator of the shell, i.e.~$u = 0.1\left(\dot{r}^2_\s{0}/2 +
r^2_\s{0}\dot{\phi}^2_\s{0}/2 + M/r_\s{0} \right)$. 
\vspace{12pt}

For both types of simulation, we take stationary boundary conditions with SPH
particles being continuously injected with constant density and velocity
distributions from a fixed injection radius $r_\s{0}$. We treat the BH horizon
(located at $r_\s{+}$) as an inner boundary at which particles are extracted
from the simulation. For a fair comparison with the toy model, we report here
late-time snapshots of the simulations in which the system has evolved to a
quasi-stationary state (at least in the region away from the disc). As mentioned
above, the number of particles being used in these was continuously changing,
but was around $2.5\times10^5$ at the time shown. Moreover, in order to reduce
the noise level and exploiting the axisymmetry of the system, the results
presented in the following were obtained after averaging over 24 cross-sectional
$\phi= \text{const.}$ slices of the 3D simulations.

\subsection{Example I}
\label{ex1}

Here we consider a set of parameter values for the system that might arise in
the context of collapsing stellar cores leading to long GRBs, namely
\begin{align}
a & = 0.5\,M,\label{e9.0}\\
M & = 4\,M_\odot,\label{e9.1}\\
\dot{M} & = 0.01\, M_\odot/s,\label{e9.2}\\
r_\s{0} & = 100\,M,\label{e9.3}\\
\dot{r}_\s{0} & = -1/\sqrt{50},\label{e9.4}\\
r_\s{0}\,\dot{\phi}_\s{0} & = 0.038,\label{e9.5}\\
\dot{\theta}_\s{0} & = 0. \label{e9.6}
\end{align}

\noindent Note that, for convenience, we have used standard (non-geometrized)
units to express the total accretion rate in Eq.\,\eqref{e9.2}. 

\begin{figure}
\begin{center}
\includegraphics[width=77mm]{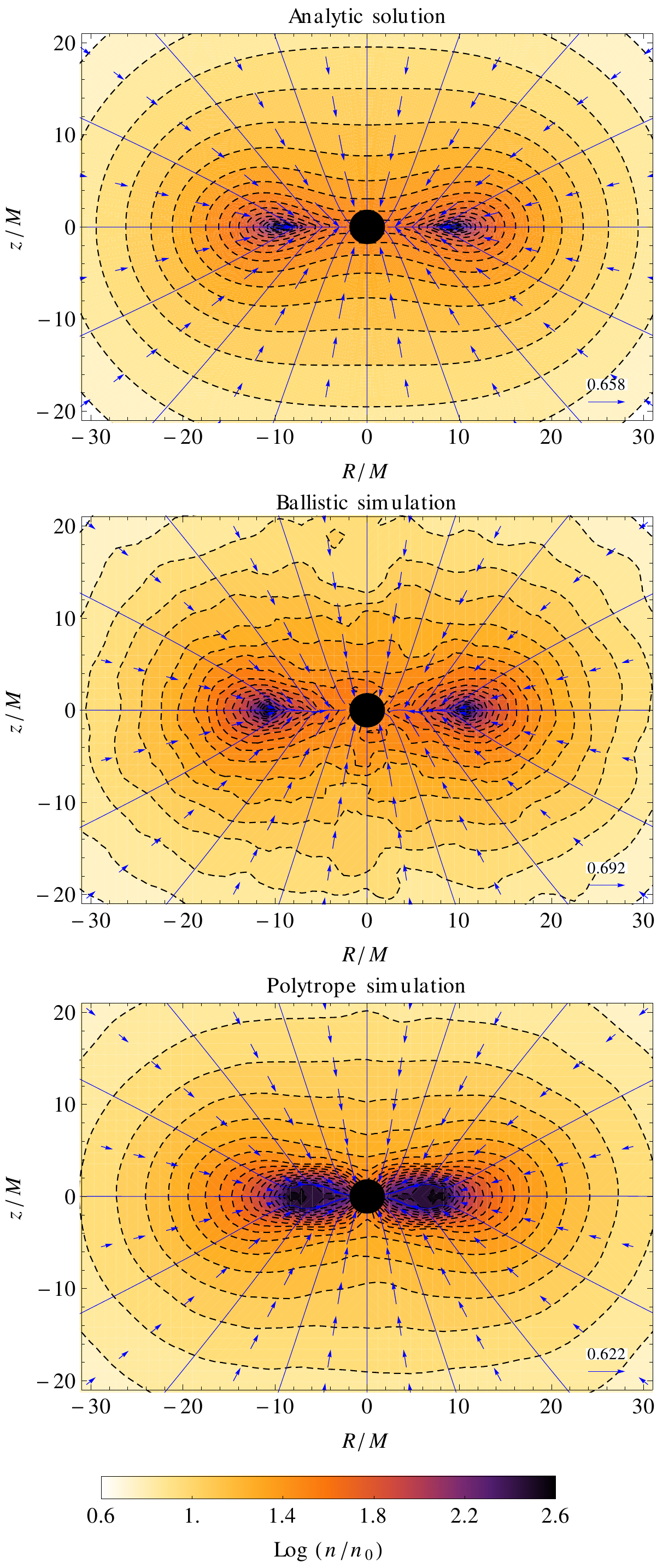}
\end{center}
 \caption{Isodensity contours, streamlines and velocity fields for the analytic
solution and for the ballistic and polytropic SPH simulations, for an accretion
flow onto a rotating BH with \hbox{$a=0.5\,M$}. The model parameters are as
given in Eqs.\,\eqref{e9.0}-\eqref{e9.6}. General relativistic effects in the
SPH simulations are mimicked using the MM pseudo-Newtonian potential. The common
scale for the density colour coding is shown at the bottom of the figure. The
velocity field in each panel is represented by the two-vectors ($V^{R}$,
$V^{z}$); the length scale for these vectors is given at the bottom right corner
of each panel. The SPH simulations used a varying total particle number, but
typically this was around $2.5 \times 10^5$ at the times shown (mass per SPH
particle $\approx 3.6\times10^{-10} M_\odot$).}
 \label{f4}
\end{figure}

For this set of boundary conditions, we present in Figure~\ref{f4} the analytic
solution alongside the results of both the ballistic and polytropic simulations.
The figure shows a spatial projection of each case onto the \hbox{$R$-$z$} plane
with isodensity contours, streamlines and velocity fields. Let us focus first on
the ballistic simulation result (middle panel), which rapidly reached a
stationary state. In this figure we see an overall satisfactory agreement with
the analytic solution, although a closer inspection of the streamlines reveals
some quantitative differences. We also note that the simulation isodensity
contours are somewhat `noisy' compared with the analytic results. Nevertheless,
this level of fluctuation is consistent with the effects of discretisation and
interpolation within SPH simulations.

\begin{figure}
\begin{center}
\includegraphics[width=84mm]{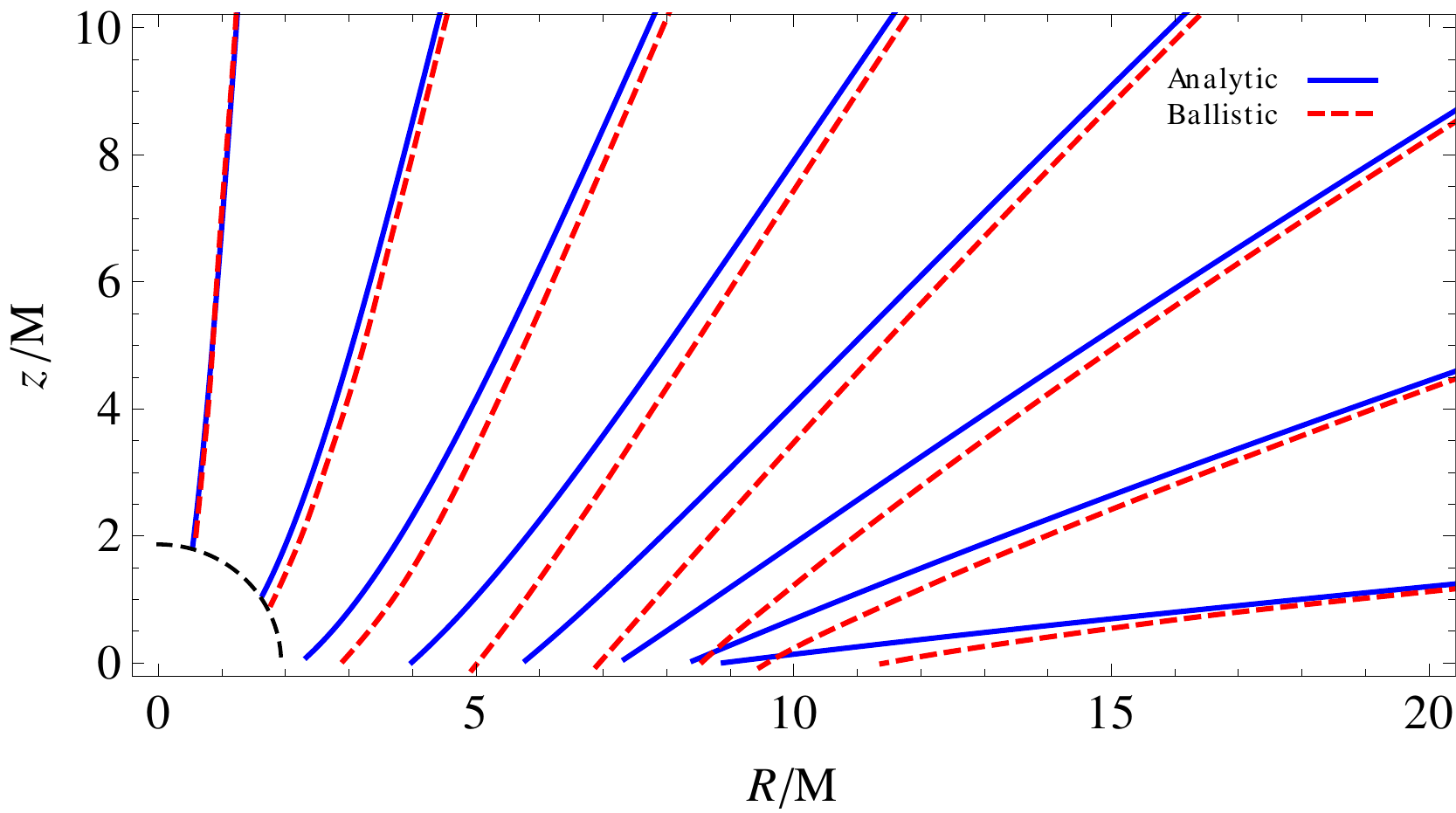}
\end{center}
 \caption{Streamlines corresponding to the analytic solution and to the
ballistic SPH simulation for an accretion flow onto a rotating BH with
\hbox{$a=0.5\,M$} (see top and middle panels of Figure~\ref{f4}). General
relativistic effects in the SPH simulation are here mimicked using the MM
pseudo-Newtonian potential. The figure shows the first quadrant of the $R$-$z$
plane with the BH horizon (located at $r_\s{+}$) indicated with a dashed-line
quarter-circle.}
 \label{f5}
\end{figure}

Figure~\ref{f5} shows a closer comparison of the streamlines of the analytic
solution with the ones of the ballistic simulation. Since hydrodynamical effects
are absent in this case, the differences between the numerical simulation and
the toy model can reasonably be attributed mainly to the different descriptions
of gravity: Kerr spacetime against the MM pseudo-Newtonian potential. From this
figure we see that the streamlines in the two cases deviate significantly from
each other just for $r\lesssim 10\,M$ and that far away from the central BH the
differences between these two descriptions of gravity become negligible (bear in
mind that the streamlines originate from $r=100\,M$).

\begin{figure}
\begin{center}
\includegraphics[width=84mm]{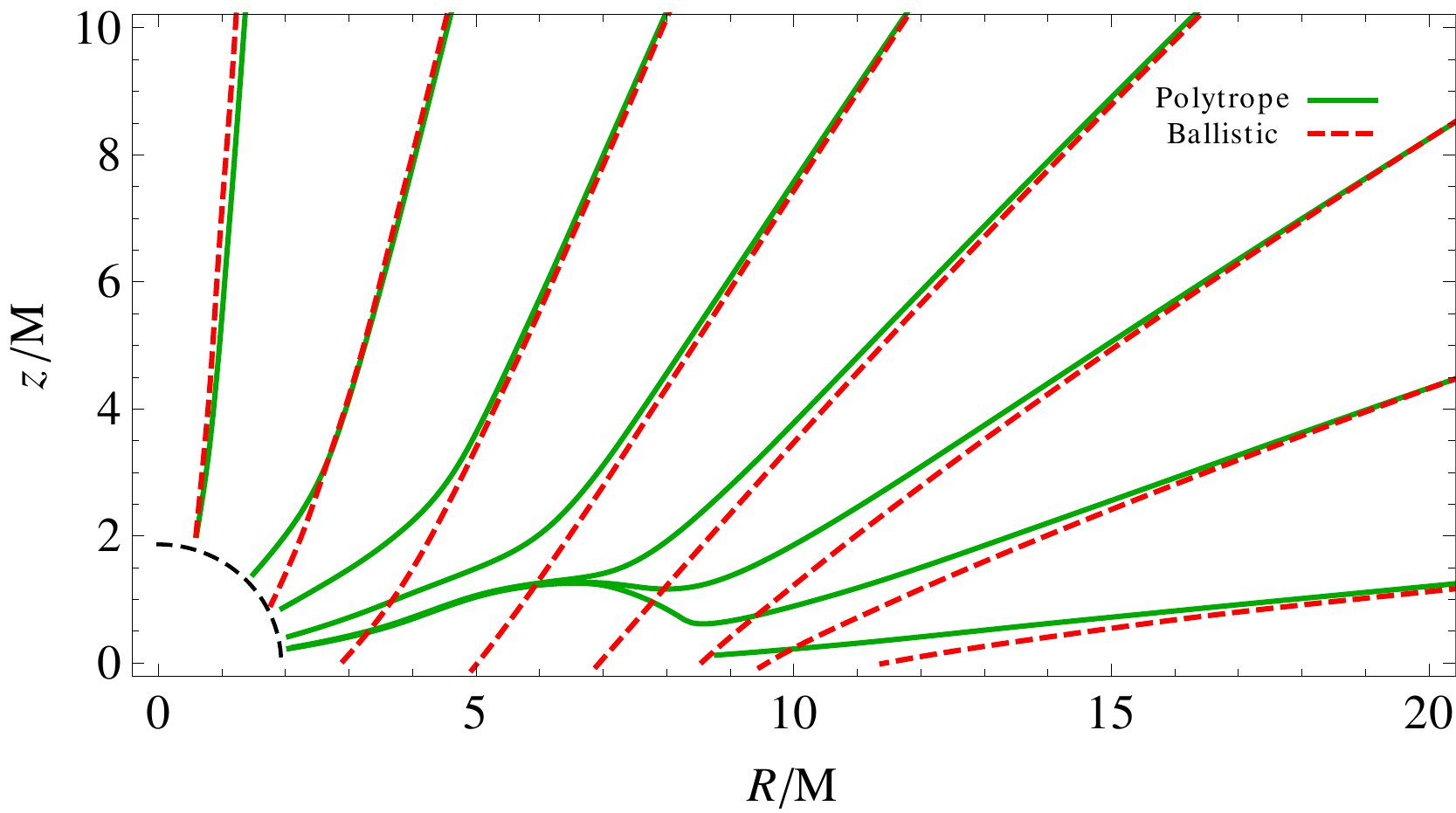}
\end{center}
 \caption{Streamlines corresponding to the ballistic and polytropic SPH
simulations made with the MM pseudo-Newtonian potential for an accretion flow
onto a rotating BH with \hbox{$a=0.5\,M$} (see the middle and bottom panels of
Figure~\ref{f4}). The figure shows a zoom-in of the first quadrant of the
$R$-$z$ plane. The BH horizon (located at $r_+$) is indicated with the
dashed-line quarter-circle.}
 \label{f6}
\end{figure}

We now investigate the inclusion of hydrodynamic properties of the flow by
comparing the polytropic SPH simulation shown in the bottom panel of
Figure~\ref{f4} with the corresponding ballistic flow shown in the middle one.
In Figure~\ref{f6} we present a direct comparison of the streamlines for these
two cases. The first important thing to note is that in the polytropic
simulation, unlike in the ballistic one, SPH particles are not removed from the
simulation at the equatorial plane and, therefore, the build-up of a disc can
take place. This is indeed the case for the present set of parameter values, for
which we observe a disc that keeps growing in mass and expands horizontally. The
material in this disc corresponds to the fraction of the infalling matter which
possesses enough angular momentum to remain in a stable orbit around the BH.
Since it is not within the scope of the present article to study the evolution
of such a disc, we show here a snapshot at a time at which any kind of initial
transient related to the initial conditions has faded away, but, at the same
time, neither the mass nor the extension of the disc have grown importantly
(additionally, the presence of cooling in the simulation aids in limiting the
disc height). 

A second important feature characterising the polytropic simulation is the
existence of a shock front around the disc, marking the boundary between two
different flow regimes. In the pre-shock region, a clear stationary regime
is rapidly reached where the flow moves supersonically and is highly laminar. In
this region, which we shall refer to as the infall region, we find that the
polytropic simulation produces quite similar results to the ballistic one, and
hence also to the toy model. On the other hand, in the post-shock region the
flow is decelerated, the streamlines deviate away from the ballistic
trajectories due to the action of pressure gradients and, in this way, are
prevented from having a `head-on' collision with their symmetric counterparts
coming from the opposite hemisphere. Clearly, the full hydrodynamical evolution
in this region will depend on the particular EoS being used as well as on the
particular mechanism driving the accretion (e.g. viscosity, dynamical
instabilities, etc.) and also the cooling prescription, but again, the details
of this post-shock region were outside the scope of the present study. From
Figure~\ref{f4} we also note that the isodensity contours of the polytropic
simulation in the infall region are less noisy than those of the ballistic
simulation, due to the action of pressure forces in smoothing out the particle
distribution and so reducing discretisation fluctuations.

Note that, in comparing Figures~\ref{f5} and \ref{f6}, the departure of the
ballistic streamlines from the analytic solution occurs earlier and for a larger
fraction of the simulation domain than the differences between the ballistic and
polytropic streamlines. In other words we see that here, adopting an improved
description for the gravitational field of the BH has a greater effect on the
infall part of the simulation than including pressure.

\subsection{Example II}

In order to make a connection with the results presented in Paper I and to
demonstrate the application of the toy model as a useful tool for studying the
effect of different gravity treatments, we considered the same boundary
conditions as in Eqs.\,\eqref{e9.1}-\eqref{e9.6} but now with $a=0$. This same
set of parameters was also used in Paper I to make a comparison with one of the
GRB simulations of \cite{lee06} using the PW potential. 

\begin{figure}
\begin{center}
\includegraphics[width=84mm]{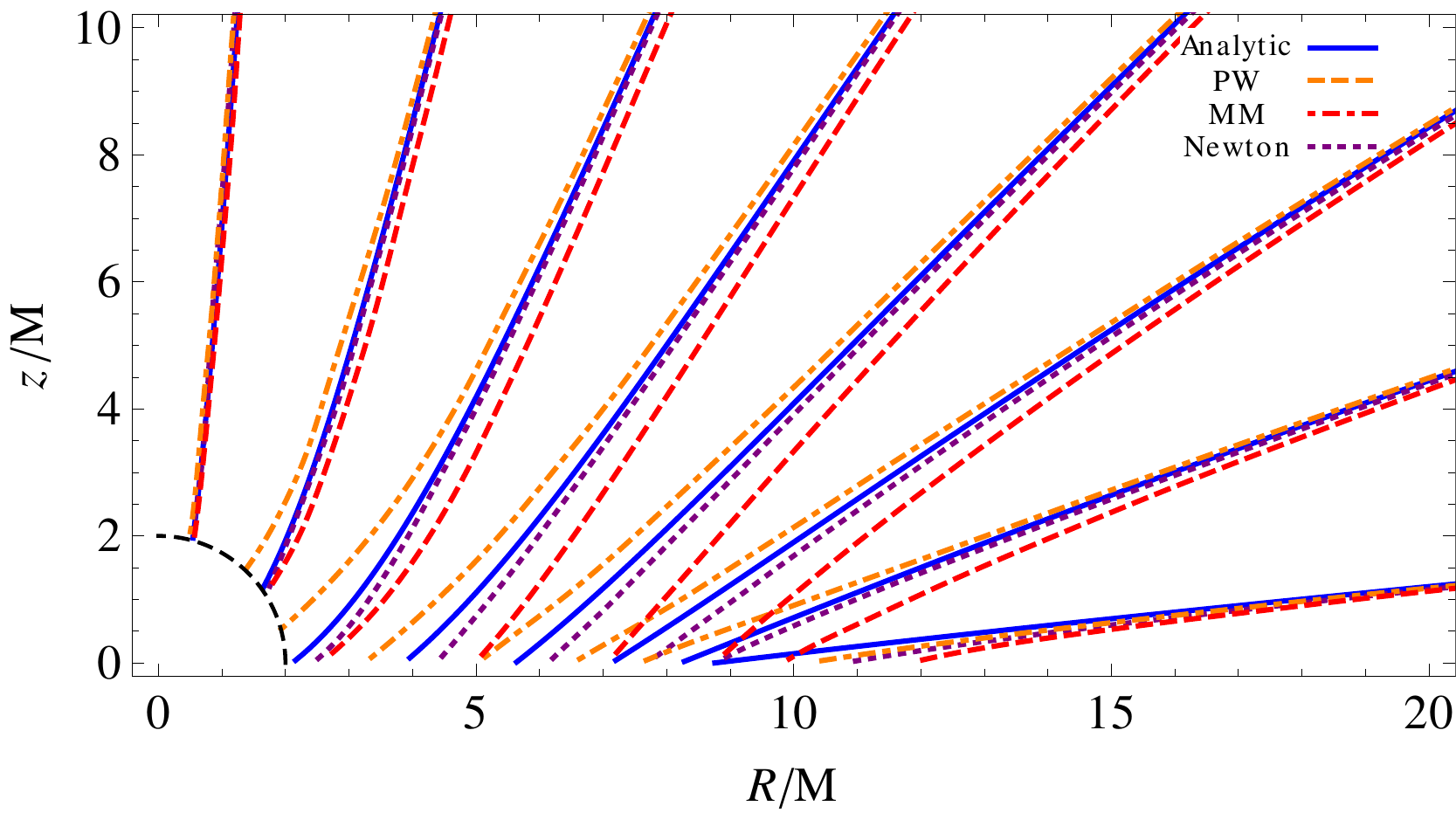}
\end{center}
 \caption{Streamlines corresponding to the analytic solution and to the three
ballistic SPH simulations for a non-rotating BH ($a=0$) with the remaining
boundary conditions being as in Eqs.\,\eqref{e9.1}-\eqref{e9.6}. The figure
shows a zoom-in of the first quadrant of the $R$-$z$ plane. The BH horizon
(located at $r_+$ and which represents the inner boundary) is indicated with the
dashed-line quarter-circle.}
 \label{f7}
\end{figure}

\begin{figure}
\begin{center}
\includegraphics[width=82mm]{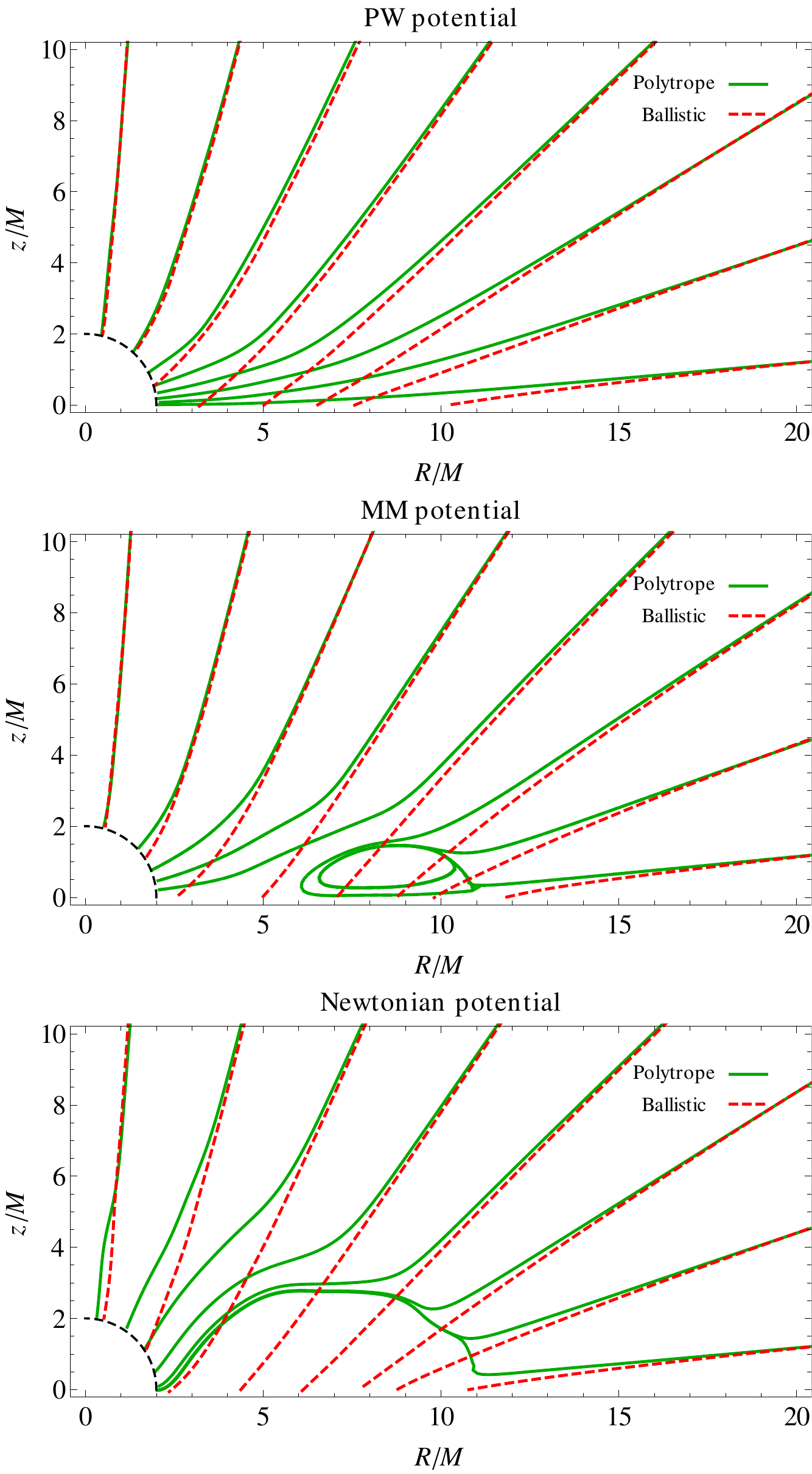}
\end{center}
 \caption{Comparison of the streamlines from SPH simulations for
ballistic motion against those for a polytropic fluid. Note that only for the
run using the PW potential does the resulting flow correspond to a `small-scale
inviscid disc'. In the other two cases, the infalling matter keeps accumulating
in a ring around the BH. This effect is more evident in the case with the MM
potential.}
 \label{f8}
\end{figure}

Figure~\ref{f7} shows a comparison of the analytic general relativistic
streamlines with those coming from ballistic SPH simulations with the three
different gravitational potentials: the classical Newtonian one, and the PW and
MM pseudo-Newtonian ones. Note how the streamlines obtained with the Newtonian
potential are closer to the general relativity solution than those obtained with
the pseudo-Newtonian potentials. In this figure we can also see that the PW
streamlines arrive at the equatorial plane at smaller radii than the analytic
relativistic ones, while the Newtonian and MM ones arrive at larger radii. This
suggests that equivalent hydrodynamical simulations implementing the PW
potential would underestimate the extension of any resulting disc while those
implementing the Newtonian and MM potentials would overestimate it.

Once more we analyse the role of pressure gradients on the infall by showing in
Figure~\ref{f8} the results from polytropic simulations for the three potentials
employed in Figure~\ref{f7} in comparison with the equivalent ballistic ones. We
see a quite good match between the two sets of streamlines in the infall region
while the effects of the pressure gradients become significant only in the high
density region near to the equatorial plane. Again, in this region we observe
that a shock front develops around the disc where the incoming streamlines
decelerate and deviate from the corresponding ballistic trajectories.

With respect to the $a\ne0$ case in Section \ref{ex1}, the change in the spin
parameter of the BH does not lead to significant differences in either the
velocity field or the density field of the accretion flow in the infall region.
Nevertheless, a crucial difference comes at the level of the long term evolution
of the resulting disc since, for the present boundary conditions with $a=0$, one
expects that none of the matter in the equatorial plane would possess enough
angular momentum to maintain a stable orbit around the central BH and, hence,
that all of the infalling material should be accreted into the BH on a dynamical
time-scale. This kind of accretion corresponds to the `small-scale inviscid
disc' regime discussed by \cite{beloborodov}. In this respect, an important
difference among the accretion flows corresponding to different 
potentials is already apparent in Figure~\ref{f8}. In this figure we see that
only the disc corresponding to the PW potential evolves as an accretion disc
within the `small-scale inviscid disc' regime (in which a stationary state is
rapidly reached). For the other two cases we observe a growing ring of matter
with enough angular momentum to avoid direct accretion onto the BH (even though
the ballistic streamlines for the Newtonian potential appear to give the best
match to those of the analytic relativistic solution in Figure~\ref{f7}); the
resulting disc in each of these cases is then expected to evolve on a viscous
time-scale rather than the much shorter dynamical time-scale.

\setcounter{equation}{0}
\section{Discussion and conclusions}
\label{discussion}

In this paper, we have presented an analytic toy model for the relativistic
accretion of non-interacting particles onto a Kerr BH. Taking the assumptions of
stationarity, axisymmetry and ballistic motion, we have given analytic
expressions for the streamlines and the velocity fields as well as a simple
numerical scheme for calculating the density field. This model is a
generalisation of the one presented in \hbox{Paper I} for Schwarzschild
spacetime, and it has been demonstrated how the earlier results are easily
recovered from the present solution in the non-rotating limit. 

Using a single analytic expression for describing the streamlines constitutes a
novel way of expressing the solution to the latitudinal and radial motion of
timelike geodesics in Kerr spacetime. The generality of this expression is shown
in the Appendix by using standard identities of the Jacobi elliptic functions. 

The conditions for the initial profiles of Eqs.\,\eqref{e1.3}-\eqref{e1.6} were
differentiability and axisymmetry with respect to the equatorial plane. While
these are very broad and applicable to a number of astrophysical situations, a
further extension of these analytic expressions would be to include more 
asymmetric situations, such as an inhomogeneous density distribution or an
overall rotation which is not aligned with the black hole angular momentum
(occurring physically, for example, in the GRB case of a kick during an
associated supernova). This is something that we plan to investigate in future
work.

We have explored the effect of frame dragging on the resulting accretion flow
and found that an effective coupling occurs between the BH spin and the angular
momentum of the infall, leading to more extended discs if the flow is
co-rotating with the BH and smaller discs in the counter-rotating case.

Our model allows for a fairly wide range of boundary conditions to be used,
making it an ideal tool for exploring the effect of different flow parameters
(accretion rates, angular momentum and density distributions, etc.) in
applications where the approximations of steady-state and axisymmetry are
reasonable ones. These assumptions are often met in some interesting
astrophysical scenarios such as under-luminous accretion towards supermassive
BHs, wind-fed X-ray binaries and collapsars in which the accretion disc remains
thin either due to efficient cooling or because it evolves within the
`small-scale inviscid' regime. In this paper we have shown a series of
comparisons between the toy model and full-hydrodynamic, numerical simulations
for a collapsar-like setup. Rather good agreement was obtained between the
simulations and the toy model, under circumstances where one might expect to
have agreement. The main discrepancies between the resulting accretion flows in
the infall region have been shown to be related more to the different treatments
of the gravitational field produced by the BH rather than to the ballistic
description of the infall. Indeed, we observed that the effects of pressure
gradients tend to become important just in the immediate proximity of the
disc, where a shock front develops and decelerates the incoming flow. A new kind
of exploratory simulation can be envisaged in which simple but general boundary
conditions are set far away from the central object and then, by using the toy
model, transported down to the region in which pressure gradients become
dominant where a proper hydrodynamical study can then be performed. This kind of
approach would greatly reduce the spatial domain of the simulation, allowing
greater resolution and reducing the computing time. 

Given the analytic nature of the present model, it provides a very practical
tool for use in benchmarking general relativistic hydrodynamics codes and this,
indeed, was the main motivation for the present work which forms part of a
larger project for building a new general relativistic SPH code. Also, this toy
model allows simple and direct comparisons between approximate methods for
including general relativistic effects in simulations on a case-by-case basis.
We have used it here to test the performance of SPH simulations implementing two
pseudo-Newtonian potentials (MM and PW) and found good overall qualitative
agreement between the toy model and the simulations, although we have also seen
that apparently small quantitative discrepancies in the flows can eventually
lead to rather different long-term evolutions. In the purely ballistic
comparisons, we found good agreement between the trajectories coming from
different gravity descriptions in the regions far away from the BH. However, as
the test particles approach the inner region, the different trajectories start
to deviate significantly from each other. How important these discrepancies are
in practice will certainly depend on the particular application; nonetheless,
from the present results we can conclude that neither of the two
pseudo-Newtonian potentials considered here is particularly well suited for
reproducing off-equatorial motion of test particles.

\section{Acknowledgements}

The authors thank Zdenek Stuchl\'ik, Stephan Rosswog and Iv\'an Zalamea for
fruitful discussions, and Sergio Mendoza for initially motivating the study of
this problem. We would also like to thank the anonymous referee for helpful
suggestions and remarks.

\bibliographystyle{mn2e}
\bibliography{references}

\include{appendix}

\end{document}

%% file: appendix.tex
\appendix

\setcounter{section}{0}
\setcounter{equation}{0}

\renewcommand{\theequation}{A.\arabic{equation}}

\section{Radial and latitudinal motion}

In Section \ref{lines} we have given analytic solutions for the radial and
latitudinal motion of a timelike geodesic in Kerr spacetime. These expressions
are general but involve the use of complex quantities in some cases. In this
appendix we consider these special cases for both types of motion and, by means
of standard identities for Jacobi elliptic functions \citep[see
e.g.][]{cayley,abramowitz}, we rewrite the solutions when necessary in such a
way that just real quantities are involved.

\subsection{Radial solution}

Let us consider the solution to the radial integral in Eq.\,\eqref{e3.2}, i.e.
\begin{equation}
\int^{ r }_{ r_\s{a} } \frac{ \ud r' }{\sqrt{\mathcal{R}(r')}} = \Phi(r).
\label{ea1.1}
\end{equation}

\noindent The general solution to Eq.\,\eqref{ea1.1} was given in
Eq.\,\eqref{e4.2}. That expression involves the use of complex quantities when
$\mathcal{R}(r)$ has complex roots and should be handled with care when
$\varepsilon=0$. In order to give alternative expressions for these cases, we
consider the various possibilities one by one:

\paragraph*{Case I:} Four real roots
\vspace{12pt}

The labelling of the roots here proceeds as described in Section \ref{radial}.
In this case the solution given in Eq.\,\eqref{e4.2} involves only the use of
real quantities. For the sake of completeness, we reproduce it here
\begin{gather}
  \Phi(r) = \frac{2\,\cn^{-1}\left(\sqrt{\frac{(r_\s{d}
-r_\s{a})(r_\s{b}-r)}{(r_\s{b}-r_\s{a})(r_\s{d}-r)}},\,k_\s{r}\right)}
{\sqrt{\varepsilon(r_\s{a}-r_\s{c})(r_\s{d}-r_\s{b})}},
\label{ea1.2}\\
  k_\s{r} = \sqrt{\frac{(r_\s{b}-r_\s{a})(r_\s{d}-r_\s{c})}
{(r_\s{d}-r_\s{b})(r_\s{c}-r_\s{a})}}.
\label{ea1.3}
\end{gather}

\paragraph*{Case II:} Two real roots and a complex conjugate pair
\vspace{12pt}

We have that  $r_\s{a}$ and $r_\s{d}$ (with $r_\s{a}<|r_\s{d}|$) are the real
roots while $r_\s{b}$ and $r_\s{c}$ form a complex conjugate pair. We define the
following three real constants
\begin{gather}
\alpha = \text{Sign}(\varepsilon)\sqrt{(r_\s{d} - r_\s{b})(r_\s{d} - r_\s{c})},
\label{ea1.4}\\
\beta = \sqrt{(r_\s{a} - r_\s{b})(r_\s{a} - r_\s{c})},
\label{ea1.5}\\
\widetilde{k}^2_\s{r}= 
\frac{( \alpha + \beta )^2 - (r_\s{d}-r_\s{a})^2 }{ 4 \alpha\beta }.
\label{ea1.6}
\end{gather}

\noindent From the definition of $k_\s{r}$ in Eq.\,\eqref{ea1.3}, it is easy to
check the following relation 
\begin{equation}
\widetilde{k}^2_\s{r} = \frac{( 1 + k_\s{r})^2 }{4 k_\s{r} }.
\label{ea1.7}
\end{equation}

 Now consider the following identity for Jacobi elliptic functions
\begin{equation}
 \cn\left( 2\sqrt{ k }\,\varphi,\,\frac{ 1 + k }{2\sqrt{ k }} \right)  =
\frac{1-k\,\sn^2(\varphi,\,k)}
{1+k\,\sn^2(\varphi,\,k)}.
\label{ea1.8}
\end{equation}

\noindent Defining $u=\cn(\varphi,\,k)$ and inverting Eq.\,\eqref{ea1.8} gives
\begin{equation}
 \cn^{-1}(u,\,k) = \frac{1}{2\sqrt{ k }}\,
\cn^{-1}\left[\frac{1-k(1-u^2)}{1+k(1-u^2)},\,\frac{1 + k}{2\sqrt{k}}\right],
\label{ea1.9}
\end{equation}

\noindent from where, and after some algebra, we can rewrite $\Phi(r)$ in
Eq.\,\eqref{ea1.2} as 
\begin{equation}
\Phi(r) = \frac{ 1 }{ \sqrt{ \varepsilon\alpha\beta } }\cn^{-1}
\left[\frac{ \beta\,r_\s{d} -\alpha\,r_\s{a} + (\alpha-\beta)r}
{ \beta\,r_\s{d} +\alpha\,r_\s{a} -
(\alpha+\beta)r},\,\widetilde{k}_\s{r}\right] ,
\label{ea1.10} 
\end{equation}
which is now an explicit real function of $r$.

\paragraph*{Case III:} Two pairs of complex conjugates 
\vspace{12pt}

This case is a new possibility for Kerr spacetime that is not present in the 
Schwarzschild case since there one of the roots of $\mathcal{R}(r)$ is always
zero and, therefore, there is at least one other real root (see Paper I). 

We start by noting that, since $\Phi(r)$ is defined in Eq.\,\eqref{ea1.1} in
terms of an integral with a complex number as its lower limit, $\Phi(r)$  is
itself a complex function of $r$. In the following, we first split $\Phi(r)$
into its real and imaginary parts and then show that the latter is independent
of $r$, and so $\Phi(r)-\Phi(r_\s{0})$ will always be a real function of $r$.

According to the labelling of the roots in Section \ref{radial}, in this
case we have $r_\s{a} = r_\s{d}^*$ and $r_\s{b} = r_\s{c}^*$, with
Re$(r_\s{a})<$Re$(r_\s{b})$. Also, it is simple to check that $\alpha$ and
$\beta$, as defined in Eq.\,\eqref{ea1.4}, now form a complex conjugate pair,
i.e. $\alpha=\beta^*$. We then introduce the following real constants
\begin{gather}
 \mu = \frac{r_\s{a}+r_\s{d}}{2},\quad
\nu=\frac{r_\s{a}-r_\s{d}}{2\,i}, \nonumber\\
 \zeta = \frac{\alpha + \beta}{2},\quad  \eta = \frac{\alpha - \beta}{2\,i}.
\label{ea1.11}
\end{gather}

\noindent Using these definitions, it is simple to check that
$\widetilde{k}_\s{r}$, as defined in Eq.\,\eqref{ea1.6}, can also be expressed
as
\begin{equation}
\widetilde{k}^2_\s{r} = \frac{\zeta^2 +\nu^2}{ \zeta^2+\eta^2 },
\label{ea1.12}
\end{equation}

\noindent from which it is clear that $\widetilde{k}_\s{r}$ is still a real
quantity. On the other hand, from Eq.\,\eqref{ea1.10} and the definitions in
Eq.\,\eqref{ea1.11}, it is simple to check that 
\begin{equation}
\frac{ \beta\,r_\s{d} -\alpha\,r_\s{a} + (\alpha-\beta)r}
{ \beta\,r_\s{d} +\alpha\,r_\s{a} -(\alpha+\beta)r} =
i\,\frac{\zeta\,\nu - \eta(r-\mu)}{\eta\,\nu + \zeta(r-\mu)},
\label{ea1.13} 
\end{equation}

\noindent i.e. the argument of the function $\cn^{-1}$ in Eq.\,\eqref{ea1.10}
is a pure imaginary number. Moreover, from Eq.\,\eqref{ea1.12} it follows that 
\begin{equation}
\sqrt{ \varepsilon\alpha\beta }\,\widetilde{k}_\s{r} =
\sqrt{ \varepsilon\left(\zeta^2+\nu^2\right)} .
\label{ea1.14} 
\end{equation}

We now consider the following identity for Jacobi elliptic functions
\begin{equation}
 \cn(\phi,\,k) = i\,\cnsn\left[k\,\phi+i\,K\left( 1+\frac{1}{k}\right) ,\,
\frac{1}{k}\right],
\label{ea1.15}
\end{equation}

\noindent where $K$ is the complete elliptic integral of the first kind. Again
defining $u=\cn(\varphi,\,k)$ and solving for $\varphi$ in Eq.\,\eqref{ea1.15}
gives
\begin{equation}
 \cn^{-1}(u,\,k) =
\frac{1}{k}\left[\cnsn^{-1}\left(-i\,u,\,\frac{1}{k} \right)-
i\,K\left( 1+\frac{1}{k}\right)\right] .
\label{ea1.16}
\end{equation}

Introducing $\widetilde\Phi(r)$ as the following real function of $r$
\begin{equation}
\widetilde\Phi(r) = \frac{ 1 }{\sqrt{\varepsilon\left(\zeta^2+\nu^2\right)}}\,
\cnsn^{-1} \left[\frac{\zeta\,\nu - \eta(r-\mu)}{\eta\,\nu + \zeta(r-\mu)},\,
\frac{1}{\widetilde{k}_\s{r}}\right],
\label{ea1.17}
\end{equation}

\noindent and combining Eqs.\,\eqref{ea1.13}, \eqref{ea1.14}, \eqref{ea1.16} and
\eqref{ea1.17}, it is simple to check that 
\begin{equation}
\Phi(r) = \widetilde\Phi(r) - \frac{i\,K\left( 1+\frac{1}{k}\right)}
{\sqrt{\varepsilon\left(\zeta^2+\nu^2\right)}}.
\label{ea1.18}
\end{equation}

\noindent This last equation explicitly splits $\Phi(r)$ into a real function of
$r$ and an imaginary constant. From this, it follows that  
\begin{equation}
\Phi(r) - \Phi(r_\s{0}) = \widetilde\Phi(r) - \widetilde\Phi(r_\s{0})
\label{ea1.19}
\end{equation}

\noindent is a real function of $r$.

\paragraph*{Case IV:}  $\varepsilon = 0$
\vspace{12pt}

In this case we have that one of the roots diverges to infinity and so
$\mathcal{R}(r)$ reduces to a third order polynomial. Here there are two
possibilities for the roots: either the three of them are real, or one is real
and the other two form a complex conjugate pair. Appropriate expressions for
each case are straightforward to obtain from Eqs.\,\eqref{ea1.2} and
\eqref{ea1.10} after taking the corresponding limit. See Paper I for an
analogous procedure.

\subsection{Polar solution}

We now return to the polar integration discussed in Section \ref{polar} and 
consider the polar solution as given in Eq.\,\eqref{e5.3}. That expression is
always a real function of $\theta$ but, for some values of $Q$ and
$\varepsilon$, it might involve the use of complex quantities as intermediate
steps. Here we show how to rewrite $\Psi(\theta)$ in a way which involves only
real quantities. For doing this, we  consider the following cases:

\paragraph*{Case I:} $\varepsilon \le 0$
\vspace{12pt}

In this case, it follows from Eq.\,\eqref{e5.2} that $Q\ge0$ and so all of
the quantities involved in Eq.\,\eqref{e5.3} are real. For the sake of
completeness, we reproduce the expression here 
\begin{equation}
\Psi(\theta) = \frac{\cos\theta_\s{a}}{\sqrt{Q}}\,\cd^{-1}
\left(\frac{\cos\theta}{ \cos\theta_\s{a}},\, k_\s{\theta}\right),
\label{ea2.1}
\end{equation} 

\noindent where
\begin{equation}
k_\s{\theta} = \sqrt{- \varepsilon\,a^2/Q}\, \cos^2\theta_\s{a}.
\label{ea2.2}
\end{equation}

Note that when $\varepsilon=0$ then $k_\s{\theta}=0$, and since
$\cd(\varphi,0)=\cos(\varphi)$ then Eq.\,\eqref{ea2.1} can be simplified as
\begin{equation}
\Psi(\theta) = \frac{\cos\theta_\s{a}}{\sqrt{Q}}\,\cos^{-1}
\left(\frac{\cos\theta}{ \cos\theta_\s{a}}\right).
\label{ea2.3}
\end{equation} 

As noted in Section \ref{polar}, this result also follows when $a=0$; hence
Eq.\,\eqref{ea2.3} is the expression to use in Schwarzschild spacetime (see
Paper I). 

\paragraph*{Case II:} $\varepsilon > 0$ and $Q>0$
\vspace{12pt}

Here one has that $k_\s{\theta}$ is a pure imaginary number. Using the
following identity for Jacobi elliptic functions
\begin{equation}
\cd(\varphi,\,k) 
=\cn\left(\frac{\varphi}{\sqrt{1-k^2}},\,\sqrt{\frac{-k^2}{1-k^2}}\right),
\label{ea2.4}
\end{equation}

\noindent one can rewrite $\Psi(\theta)$ in Eq.\,\eqref{ea2.1} as
\begin{equation}
 \Psi(\theta) =
\frac{\cos\theta_\s{a}}
{\sqrt{Q+\varepsilon\,a^2\cos^4\theta_\s{a}}}\, \cn^{-1}
\left(\frac{\cos\theta}{\cos\theta_\s{a}},\,\widetilde{k}_\s{\theta}\right),
\label{ea2.5}
\end{equation}
with
\begin{equation}
\widetilde{k}_\s{\theta} = \sqrt{\frac{-k^2_\s{\theta}}{1-k^2_\s{\theta}}} =
\sqrt{\frac{\varepsilon\,a^2\cos^4\theta_\s{a}}
{Q+\varepsilon\,a^2\cos^4\theta_\s{a}}}\, .
\label{ea2.6}
\end{equation}

\paragraph*{Case III:} $\varepsilon > 0$ and $Q\le0$
\vspace{12pt}

In this case, we use the identity
\begin{equation}
\cn(\phi,\,k) =\dn\left( k\,\phi,\,k^{-1}\right)
\label{ea2.7}
\end{equation}

\noindent to transform $\Psi(\theta)$ as written in Eq.\,\eqref{ea2.5} into
\begin{equation}
\Psi(\theta) = 
\frac{1}{\sqrt{\varepsilon}\,a\,\cos\theta_\s{a}}\,
\dn^{-1}\left(\frac{\cos\theta}{\cos\theta_\s{a}},\,
\frac{1}{\widetilde{k}_\s{\theta}} \right).
\label{ea2.8}
\end{equation}

When $Q=0$ then $\widetilde{k}_\s{\theta}=1$ and, since $\dn(\varphi,1) =
\sech(\varphi)$, Eq.\,\eqref{ea2.8} can be simplified as
\begin{equation}
\Psi(\theta) = \frac{1}{\sqrt{\varepsilon}\,a\,\cos\theta_\s{a}}\,
\sech^{-1}\left(\frac{\cos\theta}{ \cos\theta_\s{a}}\right).
\label{ea2.9}
\end{equation} 

\paragraph*{Case IV:}  $\ell = 0$
\vspace{12pt}

In this case, the expressions in Eqs.\,\eqref{ea2.1}, \eqref{ea2.5} and
\eqref{ea2.8} can be used without further modification. Nevertheless, note that
here one has the possibility of reaching the polar axis where the polar
coordinate is singular. This happens when $\ell=0$ and $Q>-\varepsilon a^2$, and
in this case one should take $\theta_\s{a} = 0$ which, although it is not a
formal root of the equation $\Theta(\theta) = 0$, does constitute a turning
point of the polar motion since here one has that the polar velocity changes
sign discontinuously every time that the particle crosses the polar axis.

\label{lastpage}